\documentclass[sigconf]{acmart}
\AtBeginDocument{%
  \providecommand\BibTeX{{%
    \normalfont B\kern-0.5em{\scshape i\kern-0.25em b}\kern-0.8em\TeX}}}

\copyrightyear{2023} 
\acmYear{2023} 
\setcopyright{acmlicensed}\acmConference[CHI '23]{Proceedings of the 2023 CHI Conference on Human Factors in Computing Systems}{April 23--28, 2023}{Hamburg, Germany}
\acmBooktitle{Proceedings of the 2023 CHI Conference on Human Factors in Computing Systems (CHI '23), April 23--28, 2023, Hamburg, Germany}
\acmPrice{15.00}
\acmDOI{10.1145/3544548.3580741}
\acmISBN{978-1-4503-9421-5/23/04}
\newcommand{\tool}{\textsc{DeepLens}}

\newcommand{\circled}[1]{{\large \textcircled{\footnotesize #1}}}

\usepackage{xcolor}

\definecolor{primary}{HTML}{3f51b5}
\definecolor{secondary}{HTML}{f44336}

\definecolor{lightgray}{HTML}{eeeeee}
\definecolor{tab_red}{rgb}{1,0.76,0.71}
\definecolor{tab_purple}{HTML}{FDD0D0}

\usepackage{multirow}

\usepackage{xspace}
\usepackage[ruled,linesnumbered]{algorithm2e}    
\usepackage{colortbl}
\usepackage{enumitem}

\makeatletter
\DeclareRobustCommand\onedot{\futurelet\@let@token\@onedot}
\def\@onedot{\ifx\@let@token.\else.\null\fi\xspace}

\def\eg{e.g\onedot}

\def\etal{et al\onedot}
\makeatother

\newcommand{\responseref}[0]{\color{black}}
\newcommand{\responseline}[1]{\textcolor{black}{#1}}
\newcommand{\camerareadyrevision}[1]{\textcolor{black}{#1}}

\usepackage{acmart-taps}
\begin{document}

\title{\tool : Interactive Out-of-distribution Data Detection in NLP Models}

\author{Da Song}
\authornote{Both authors contributed equally to this work.}
\affiliation{%
  \institution{University of Alberta}
  \city{Edmonton}
  \state{AB}
  \country{Canada}}
\email{dsong4@ualberta.ca}

\author{Zhijie Wang}
\authornotemark[1]
\affiliation{%
  \institution{University of Alberta}
  \city{Edmonton}
  \state{AB}
  \country{Canada}
}
\email{zhijie.wang@ualberta.ca}

\author{Yuheng Huang}
\affiliation{%
  \institution{University of Alberta}
  \city{Edmonton}
  \state{AB}
  \country{Canada}}
\email{yuheng18@ualberta.ca}

\author{Lei Ma}
\affiliation{%
  \institution{University of Alberta, Canada}
  \institution{The University of Tokyo, Japan}
  \city{}
  \state{}
  \country{}
  }
\email{ma.lei@acm.org}

\author{Tianyi Zhang}
\affiliation{%
  \institution{Purdue University}
  \city{West Lafayette}
  \state{IN}
  \country{USA}}
\email{tianyi@purdue.edu}

\begin{abstract}

Machine Learning (ML) has been widely used in Natural Language Processing (NLP) applications. A fundamental assumption in ML is that training data and real-world data should follow a similar distribution. However, a deployed ML model may suffer from out-of-distribution (OOD) issues due to distribution shifts in the real-world data. Though many algorithms have been proposed to detect OOD data from text corpora, there is still a lack of interactive tool support for ML developers. In this work, we propose {\tool}, an interactive system that helps users detect and explore OOD issues in massive text corpora. 
Users can efficiently explore different OOD types in {\tool} with the help of a text clustering method. Users can also dig into a specific text by inspecting salient words highlighted through neuron activation analysis.
In a within-subjects user study with 24 participants, participants using {\tool} were able to find nearly twice more types of OOD issues accurately with 22\% more confidence compared with a variant of {\tool} that has no interaction or visualization support.

\end{abstract}




\begin{teaserfigure}
  \centering
  \includegraphics[width=\linewidth]{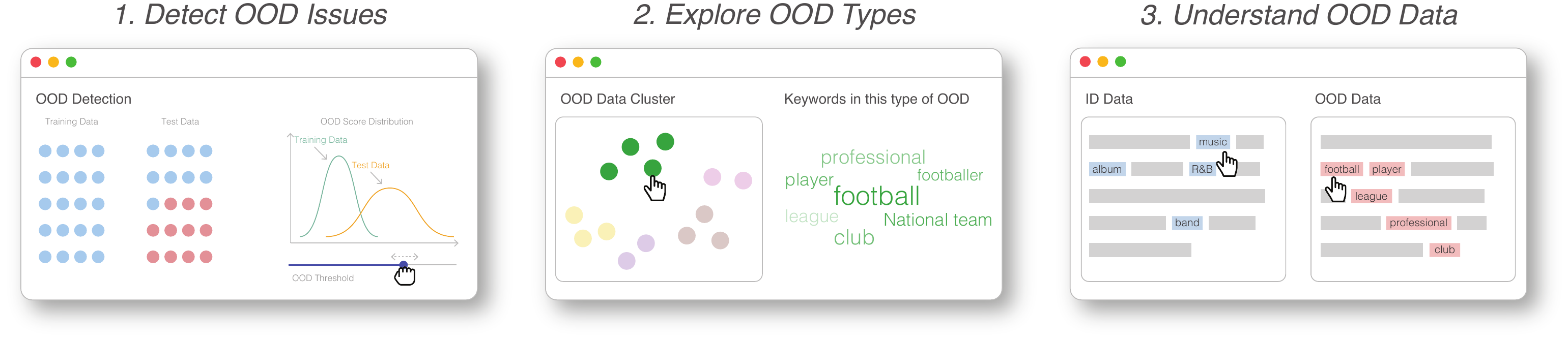}
  \caption{\textbf{{\tool} is an interactive system for supporting out-of-distribution (OOD) data detection in NLP models.} The developer can \textbf{detect OOD issues} by dynamically adjusting the threshold and observing the changes in the icon array and OOD score distribution. {\tool} also helps the developer \textbf{explore OOD types} by clustering similar texts and visualizing keywords. To \textbf{understand OOD data}, the developer can check the highlighted keywords and compare them with in-distribution (ID) data.}
  \Description{This figure shows the three core features of DeepLens which can help users detect, explore and understand OOD issues. The first feature help developers detect out-of-distribution  (OOD) issues by dynamically adjusting the OOD score threshold and observing the changes in the icon array and OOD score distribution. The second feature enables users to explore OOD types by clustering similar texts and visualizing keywords. The third feature can highlight keywords of selected data and users can compare OOD data with (in-distribution) data to understand OOD data.}
  \label{fig:teaser}
\end{teaserfigure}

\maketitle

\section{Introduction}
\label{sec:intro}

Machine Learning (ML) techniques and ML models have shown superior performance in many applications, \eg, autonomous driving~\cite{peng2020first}, virtual assistant~\cite{bocklisch2017rasa}, and medical diagnosis~\cite{li2014deep}. 
Modern ML techniques usually assume the training data and test data follow a similar distribution. However, such an assumption can hardly be satisfied in the real world. Instead, data distribution shift and out-of-distribution (OOD) samples often result in performance degradation of deployed ML models \cite{hendrycks2016baseline, louizos2017multiplicative, guo2017calibration}. Such performance drop further brings concerns about the reliability and trustworthiness of ML models. In particular, failing to predict OOD samples may lead to serious outcomes in high-stake and safety-critical applications such as health care~\cite{li2014deep}. In 2013, Google Flu Trends (GFT) model failed to predict the flu season, missing the peak of that flu season by 140 percent~\cite{lazer2014parable}. One significant factor of the failure is that GFT did not take into account how users' search behavior had changed since 2012~\cite{lazer2014parable}. In this case, the data distribution shifted in 2013 compared with 2012 and eventually led to poor model performance.


To alleviate the effects of OOD data, many techniques have been proposed for OOD detection~\cite{hendrycks2016baseline, liang2018enhancing, liu2020energy, lin2021mood, wang2021can, arora2021types, morteza2022provable}. For a given data instance, these techniques typically first calculate a specific score (OOD score), and then compare it with a pre-defined threshold to determine if the data instance is OOD. However, in practice, only identifying OOD data is not sufficient for ML developers. For instance, in the previous example of GFT, after knowing there is an OOD issue, developers still need to dig into the OOD data and figure out why they are considered OOD and what their characteristics are. This is a time-consuming process. Though a recent technique~\cite{chen2020oodanalyzer} has been proposed to address this challenge, it is only designed for image data, not text data. Compared with images, which are more glanceable for humans~\cite{grady1998neural}, more cognitive efforts are required to read and understand text data. 
Without appropriate tool support, it can be challenging and time-consuming for developers to investigate a massive amount of OOD text data at scale.

In this paper, we explore interactive tool support for helping users quickly detect and contextualize OOD samples from large text corpora. 
We present {\tool}, a novel interactive system that enables users to \textit{detect}, \textit{explore}, and \textit{understand} OOD issues. {\tool} is built upon maximum softmax probability (MSP), a popular calibration-based OOD detection method for text data~\cite{hendrycks2016baseline,arora2021types}. To help users explore different types of OOD data in text corpora, {\tool} first clusters similar OOD data by topics and then renders the frequent words in each cluster in a word cloud to help users examine and understand the topic of each cluster. As users delve into individual OOD instances, {\tool} highlights salient words in each instance via neuron activation analysis method~\cite{alammar-2021-ecco}. In this way, {\tool} helps users quickly understand a long text without reading it in detail. {\tool} also renders in-distribution and out-of-distribution data side by side to help users compare and contrast them.

To evaluate the usability and efficiency of {\tool}, we conducted a within-subjects user study with 24 programmers with various levels of expertise in ML and NLP. 
We created a comparison baseline by disabling the cluster view and the highlighting view in {\tool}. The results show that participants using {\tool} were able to find more types of OOD data on four different NLP tasks. The mean difference in the number of OOD types found by each participant using {\tool} and the baseline tool is 3.54 vs. 1.25 (Welch's $t$-test: $p<0.0001$). Participants using {\tool} also felt more confident about OOD issues they found in the ML models. The median values are 6 vs. 5 on a 7-point Likert scale (Welch's $t$-test: $p=0.002$). These results demonstrate that {\tool} can significantly improve ML developers' productivity when dealing with out-of-distribution issues in NLP models.

\responseline{In summary, the main contribution of this paper is {\tool}, an interactive system that helps users detect, explore, and understand OOD data in large text corpora. We have open-sourced our system on GitHub~\footnote{\href{https://github.com/momentum-lab-workspace/DeepLens}{https://github.com/momentum-lab-workspace/DeepLens}}. A within-subjects user study demonstrates the effectiveness of {\tool} in detecting and analyzing different types of OOD issues on a variety of NLP tasks.}

\section{Background and Related Work}
\label{sec:related_work}

\subsection{Out-of-Distribution Issues in ML Systems}

A fundamental assumption in machine learning theory is that training and test data follow a similar distribution~\cite{mohri2018foundations}. However, after model deployment, it is not uncommon to encounter real-world data that is \textit{out-of-distribution} compared with the training data. Previous studies demonstrate that when feeding OOD samples, ML models can provide erroneous predictions with high confidence~\cite{goodfellow2014explaining, nguyen2015deep}. Such errors can have serious consequences when the
predictions inform real-world decisions such as medical diagnosis, e.g. falsely classifying a healthy sample as pathogenic or vice versa~\cite{ponsero2019promises, ren2019likelihood, amodei2016concrete}.

Over the years, there has been an ongoing effort in trying to understand OOD issues in ML systems. Moreno-Torres \etal~\cite{moreno2012unifying} present a unified framework to analyze the distribution shift. Given a classification task $\mathcal{X}\rightarrow\mathcal{Y}$, the joint probability of $x\in\mathcal{X}$ and $y\in\mathcal{Y}$ can be represented as $p(y, x) = p(y|x) p(x)$. Moreno-Torres \etal~\cite{moreno2012unifying} then categorize OOD into two types: (1) {\em covariate shift}, where the input distribution $p(x)$ changes, and (2) {\em concept shift}, where the relationship between the input and class variables $p(y|x)$ changes. 
Arora \etal~\cite{arora2021types} further extend this taxonomy to NLP tasks. They assume text data can be represented as background features (e.g. genre) that are invariant across different labels, and semantic features (e.g. sentiment words) that are discriminative for the prediction
task. Therefore, they define the change of background features as \textit{background shift} and the distribution change of semantic features as \textit{semantic shift}.
In this work, we follow the OOD taxonomy and terminologies by Arora \etal~\cite{arora2021types}, since our work focuses on OOD issues in text data.

\subsection{OOD Detection}

There is a large body of literature on OOD detection in the ML community. Most of the prior work calculates an OOD score for each input, and uses a threshold to separate ID data from OOD data. Hendrycks \etal~\cite{hendrycks2016baseline} first propose a simple method to detect OOD samples, representing one of the earliest attempts in this direction. They utilize the probability of a prediction (i.e., model confidence) as the indicator for OOD issues, in which a lower probability yields a higher OOD score. However, since DL models often ``confidently'' make errors~\cite{nguyen2015deep}, only leveraging model confidence hinders further improvement of OOD detection. To address this issue, some recent work proposes to train a calibrated model, so it can give predictions with low confidence on OOD data. The calibrated model can be obtained via data augmentation~\cite{thulasidasan2019mixup, yun2019cutmix, hendrycks2020augmix}, adversarial training~\cite{bitterwolf2020certifiably, hein2019relu, choi2019novelty}, and uncertainty modelling~\cite{meinke2019towards, bibas2021single}. Another line of work to address the model confidence barrier is to leverage other indicators for OOD detection~\cite{liang2018enhancing, liu2020energy, lin2021mood, wang2021can, arora2021types, morteza2022provable}. One of the representative works is ODIN~\cite{liang2018enhancing}, which uses temperature scaling and input perturbation for OOD score computation. Furthermore, OOD detection can also be achieved by estimating the ID distribution and measuring how far the input instance is from the ID distribution~\cite{lee2018simple, kim2019guiding, hsu2020generalized}. In this work, we develop {\tool} on top of an OOD algorithm by Arora et al.~\cite{arora2021types}, which utilizes maximum softmax probability (MSP) for OOD detection in text corpora.

So far, most efforts have been put into improving the accuracy of OOD detection algorithms. However, only providing an OOD score and a list of OOD samples is insufficient for humans to understand and reason about OOD data. {\tool} fills the gap by providing an interactive system that helps developers explore OOD data detected from large text corpora and understand their characteristics.

\subsection{Interactive Support for OOD Detection}

In the past two years, there has been a growing interest in providing interactive tool support for detecting distribution change~\cite{yang2020diagnosing, chen2020oodanalyzer, yeshchenko2021visual, wang2020conceptexplorer, palmeiro2022data+, olson2021contrastive}. 
OoDAnalyzer~\cite{chen2020oodanalyzer}, an interactive system for analyzing OOD issues in image data. It provides a grid-based visualization that shows individual OOD images in a grid view. Furthermore, it allows users to zoom into individual OOD instances and highlights the parts of an image that contributes significantly to the prediction result. The main difference between OoDAnalyzer and {\tool} is that OoDAnalyzer focuses on image data while {\tool} focuses on text data. Compared with text data, images are more glanceable. Thus, the interface design in OoDAnalyzer is not applicable to OOD analysis in text corpora. To fill the gap, {\tool} leverages a text clustering method and also highlights salient words in individual text documents to help users explore and understand OOD instances.

\sloppy Data drift detection is closely related to OOD detection. Yeshchenko \etal propose Visual Drift Detection (VDD) \cite{yeshchenko2021visual}, a visualization and interaction system for detecting and analyzing business process drift. By utilizing a set of interactive charts, VDD presents the business process drift (event sequence data) in a time-dependent way. Wang \etal present ConceptExplorer \cite{wang2020conceptexplorer}, a visual analytics system for analyzing concept drifts from multi-source time-series data.
Yang \etal propose DriftVis~\cite{yang2020diagnosing}, an visual analytics system for analyzing concept drift in streaming data. It utilizes an incremental Gaussian mixture model to detect samples with concept drift and presents prediction-level visualization that reveals the performance change of the target model. \responseline{DriftVis is specifically designed for concept drift (i.e., semantic shift in NLP), while {\tool} does not have a specialized design for a particular type of distribution shift and thus can be applied to both shift types.}

{\responseref{}
\section{USER NEEDS AND DESIGN RATIONALE}
\label{sec:design}

\begin{figure*}[t]
  \centering
  \includegraphics[width=0.95\linewidth]{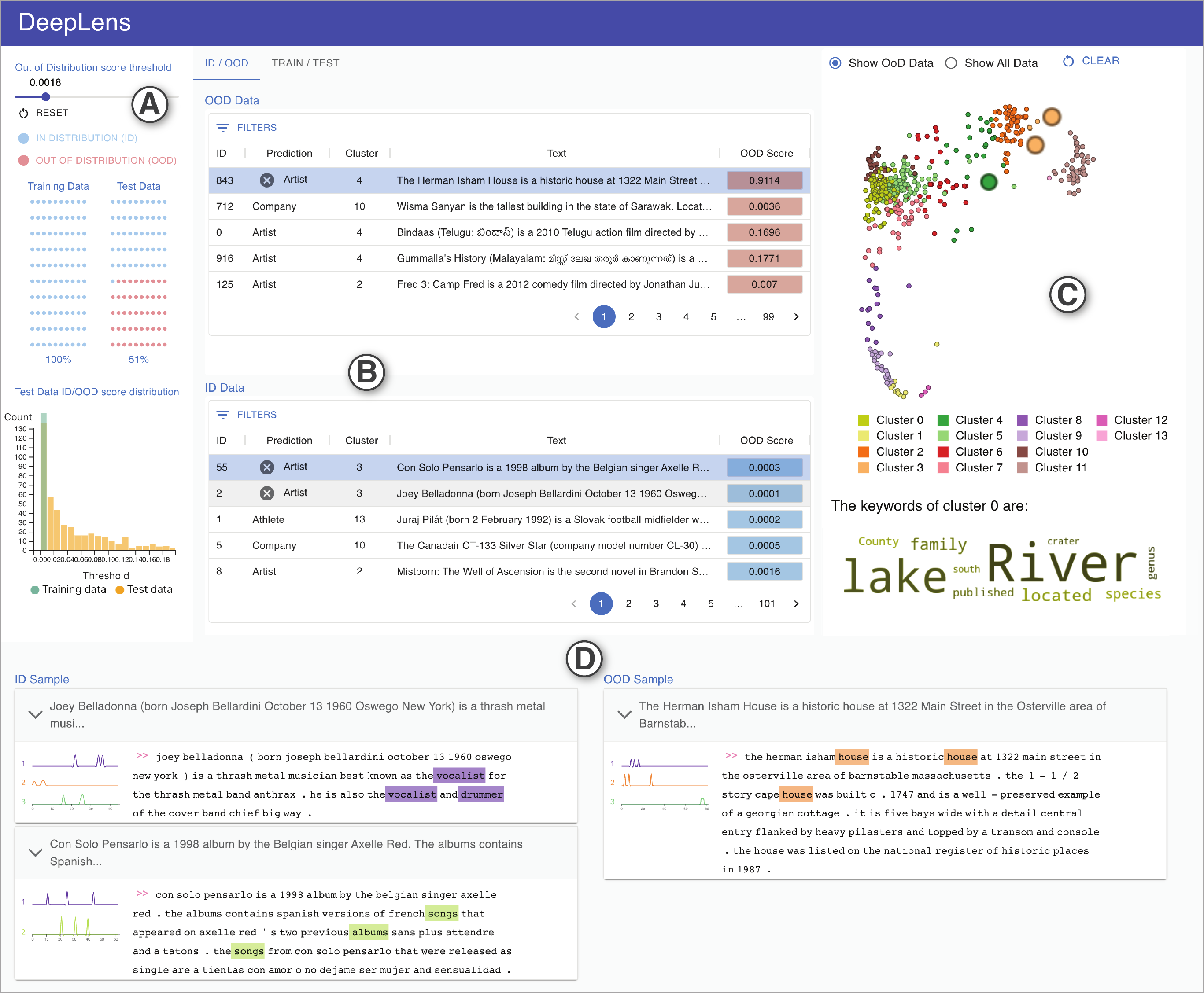}
  \caption{\textbf{\tool, an interactive system for detecting and identifying OOD samples in the text data.} \textbf{(A) The \textit{Distribution View}} allows users to adjust thresholds and inspect OOD issues in test data dynamically. \textbf{(B) The \textit{Instance View}} displays the in-distribution (ID) and out-of-distribution (OOD) data in two separate interactive data grids. \textbf{(C) The \textit{Cluster View}} displays the clustering results and keywords for each cluster for exploring potential OOD types. \textbf{(D) The \textit{Highlighting View}} shows the highlighted salient words on selected data instances to ease users' reading efforts.}
  \Description{This figure shows the interface of DeepLens, which includes four different views. (A) The first view is the Distribution View, which is at the left of the interface and contains icon arrays to show how many OOD samples might exist in the test data and a bar chart to show the OOD score distribution. The Distribution View allows users to adjust the threshold and inspect OOD issues in test data dynamically. (B) The second view is the Instance View. This view displays the ID/OOD data as two interactive tables for the user to explore the difference between ID data and OOD data. (C) The third view is the Cluster View, which locates at the right of the interface and displays the clustering results and keywords for each cluster exploring potential OOD types. (D) The fourth view is the Highlighting View, which locates at the bottom of the interface and shows the highlighted keywords on selected data instances to decrease users’ reading efforts.
}
  \label{fig:interface}
\end{figure*}

In this section, we first analyze ML practitioners' needs for interactive OOD detection based on the literature review. Then, we discuss how our proposed system supports these needs through a system overview.

\subsection{User Needs in Detecting and Diagnosing OOD Issues in ML}

To understand the needs of ML practitioners, we conduct a literature review of previous work that has done a formative study of OOD detection~\cite{chen2020oodanalyzer, yang2020diagnosing, yeshchenko2021visual, palmeiro2022data+}, has done a user study~\cite{olson2021contrastive, wang2020conceptexplorer}, or has discussed the challenges in handling OOD issues~\cite{lu2018learning, rabanser2019failing}. Based on this review, we summarize five major user needs for OOD detection.

{\noindent \textbf{\textit{N1: Automatically detect OOD data.}}}
Manually inspecting individual instances to identify data distribution shifts is time-consuming and cumbersome \cite{wang2020conceptexplorer, yeshchenko2021visual}. By working closely with their industry partners, Yeshchenko \etal~\cite{yeshchenko2021visual} found that industrial practitioners demanded the distribution shift be identified promptly and precisely. Therefore, {\tool} should automatically detect OOD data based on user-defined criteria. The expert review in Wang \etal~\cite{wang2020conceptexplorer} also confirmed the necessity of automated OOD detection in large datasets. 

\vspace{1mm}

{\noindent \textbf{\textit{N2: Understanding why a sample is detected as OOD.}}} 
Recent studies~\cite{yang2020diagnosing, wang2020conceptexplorer, lu2018learning, palmeiro2022data+} show that only detecting OOD samples is insufficient. In practice, ML practitioners are often also eager to know why those samples are out of distribution. For instance, Yang~\etal~\cite{yang2020diagnosing} interviewed four ML practitioners and found that, instead of simply obtaining the detected OOD samples, ML practitioners desired to know why and where the distribution shift occurred. In another interview study with four data scientists, Palmeiro \etal~\cite{palmeiro2022data+} found that data scientists wanted to know which parts of the dataset include data shift as well as the patterns of data shift. 

\vspace{1mm}

{\noindent \textbf{\textit{N3: Identifying different types of OOD data.}}}
Different OOD instances may have different characteristics. Therefore, ML practitioners want to categorize OOD data to better understand their commonalities and variations, so that they can come up with a more comprehensive strategy to consider the impacts brought by data shift \cite{chen2020oodanalyzer, yeshchenko2021visual, wang2020conceptexplorer, rabanser2019failing}. For example, Wang \etal~\cite{wang2020conceptexplorer} highlighted that users should be able to discriminate different types of data and verify the distribution shift of each type. In an interview study with both ML developers and ML users, Chen \etal~\cite{chen2020oodanalyzer} reported that the ML developers and ML users both desired to visually explore different types of OOD samples and their relationships to reduce the samples that need to be inspected. 

\vspace{1mm}

{\noindent \textbf{\textit{N4: Comparing OOD with ID data.}}}
Chen \etal~\cite{chen2020oodanalyzer} found that comparing OOD samples with ID samples under the same predicted label was helpful for users to confirm potential OOD issues. Olson \etal~\cite{olson2021contrastive} conducted a user study with sixty ML users and found that users often compared OOD samples with ID samples to understand the characteristics of OOD samples. 

\vspace{1mm}

{\noindent \textbf{\textit{N5: Investigating OOD issues from both global and local perspectives.}}} 
When inspecting OOD issues, users tend to first explore different categories of potential OOD data and then delve into a category of interest to compare an OOD sample with similar ID samples~\cite{chen2020oodanalyzer}. Yeshchenko \etal~\cite{yeshchenko2021visual} highlighted the importance of supporting ``drill-down'' and ``roll-up'' analysis on OOD data to allow users to flexibly investigate OOD issues from different granularity.

\subsection{Design Rationale}

To support \textbf{N1}, {\tool} leverages a calibration-based method~\cite{hendrycks2016baseline} to automatically detect OOD data in a large text corpus. Users can observe the percentage of OOD instances in the test data in the \textit{Distribution View} (Fig.~\ref{fig:interface}~\circled{A}) and adjust to what extent an instance should be considered as OOD via the threshold slider. 
To help users better understand why some instances are detected as OOD data (\textbf{N2}), \camerareadyrevision{{\tool} allows users to compare an OOD instance and an ID instance side by side and examine the commonalities and variations between them (Fig.~\ref{fig:interface}~\circled{B})}. Furthermore, as some instances are lengthy, {\tool} leverages neuron activity analysis~\cite{hewitt2019designing} to identify and highlight salient words in an instance, so users can quickly grasp the underlying topic(s) in the instance (Fig.~\ref{fig:interface}~\circled{D}). By directly comparing the differences between highlighted words in an OOD data and an ID data, users can easily identify potential topic differences between those two instances without the necessity of reading through the entire text document.

\begin{figure*}[t]
    \centering
    \includegraphics[width=0.85\linewidth]{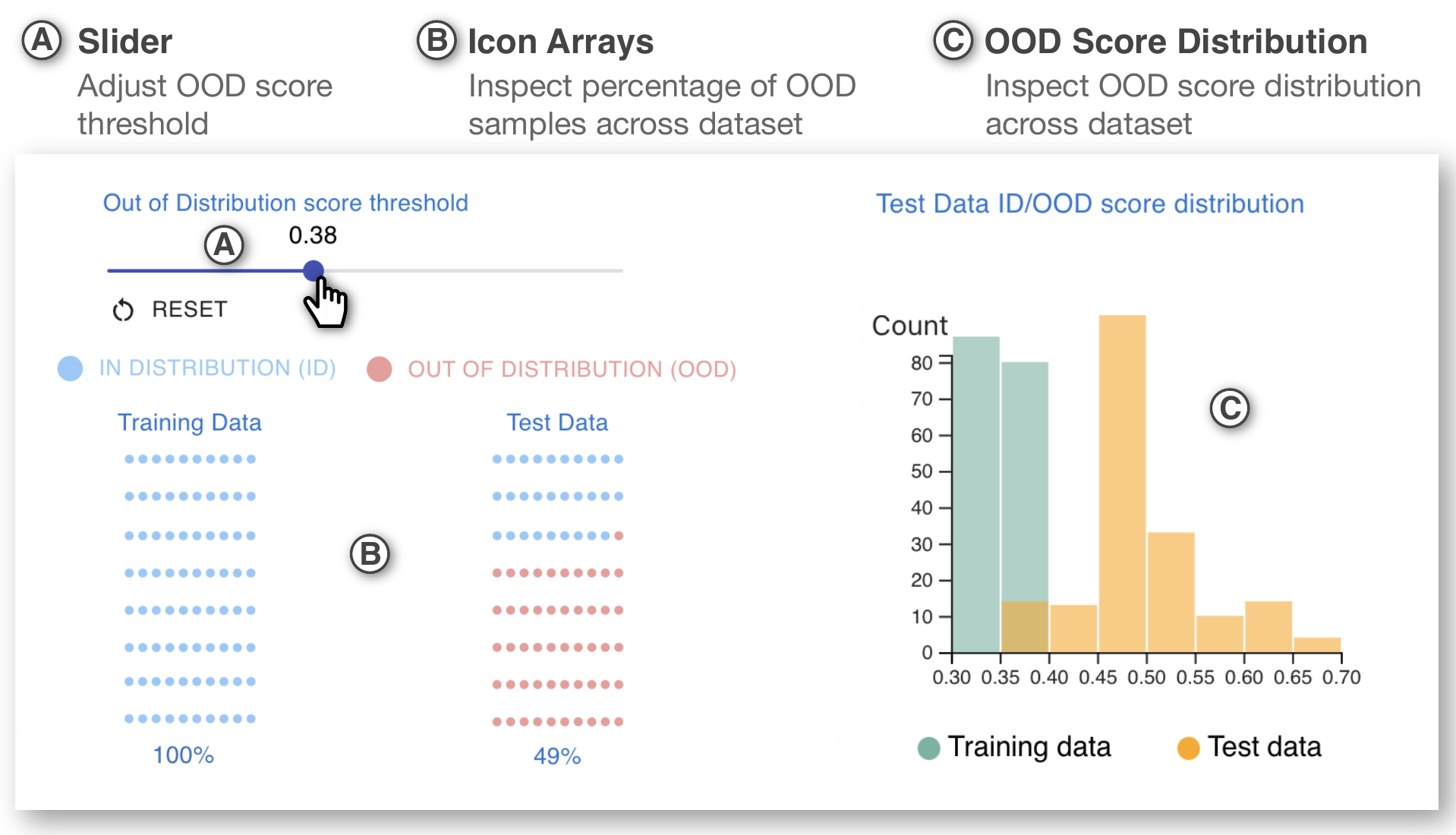}
    \caption{Users can adjust the threshold of the OOD score and inspect OOD issues in the \textbf{\textit{Distribution View}}.}
    \Description{This figure displays the Distribution View and its interaction. Users can adjust the threshold and inspect the portion of OOD samples in the test data and the OOD score distribution.}
    \label{fig:control_panel}
\end{figure*}

To assist users in identifying different types of OOD data (\textbf{N3}), {\tool} clusters the detected OOD instances and renders them in a scatter plot (Fig.~\ref{fig:interface}~\circled{C}). The common words in a cluster are visualized as a word cloud to help users understand its underlying topics (Fig.~\ref{fig:interface}~\circled{C}). This cluster view, together with the word cloud, helps users obtain a global understanding of when and where data shifts occur in the dataset (\textbf{N5}). To support the ``drill-down'' analysis mentioned in (\textbf{N5}), {\tool} allows users to delve into a specific cluster by clicking on a node in that cluster or a cluster legend. The instance view will be filtered accordingly. To support the ``roll-up'' analysis mentioned in (\textbf{N5}), {\tool} highlights the user-selected instances in the cluster view, so users can easily see where the selected instances are in the global view. Finally, to support \textbf{N4}, {\tool} allows users to filter the instances by prediction labels in the {\em Instance View} and then select OOD and ID instances with the same prediction label to compare side-by-side. The salient word highlighting feature also helps users quickly see the commonalities and variations between the OOD and ID instances.

}

\section{Design and Implementation}
\label{sec:approach}
{\noindent}In this section, we follow the 3-step usage of {\tool} (Fig.~\ref{fig:teaser}) to introduce its design and implementation: (1) detect OOD issues, (2) explore OOD types, and (3) understand OOD data. 

\subsection{Interactive OOD Text Detection}
\label{subsec:detect}
{\noindent \textbf{\textit{OOD Detection Method.}}} Given a data instance $x$, an OOD detection method first computes an OOD score $s(x)$. If $s(x) > \epsilon$ (a pre-defined threshold), then $x$ is considered as an OOD sample. {\tool} leverages a calibration-based method, \textit{MSP} (maximum softmax probability)\responseline{~\cite{hendrycks2016baseline}}, to compute the OOD score. A higher MSP means the model is highly confident with the prediction, thus a lower MSP indicates the given data instance $x$ is more likely to be an OOD sample. Given a probabilistic classifier $\mathcal{C}$,
\begin{equation}
    \label{eq:msp}
    s(x) = 1 - \max_{k}\mathcal{C}(y=k|x)
\end{equation}
\responseline{where $k\in{1, \dots, N}$ denotes class label $k$ and $y$ denotes the prediction of $C$.}
Note that a probabilistic classifier typically exists in an NLP model even if it is not for classification. For instance, a probabilistic classifier exists when projecting hidden states into the vocabulary.

\vspace{1mm}
{\noindent \textbf{\textit{Distribution View.}}} {\tool} allows users to adjust the threshold of OOD detection dynamically via adjusting the slider (Fig.~\ref{fig:control_panel}~\circled{A}). When the threshold is updated, the \textit{icon arrays} (Fig.~\ref{fig:control_panel}~\circled{B}) will also be updated accordingly. \responseline{We chose to use \textit{icon arrays} since it provides a discrete-event representation and has been proven to lead to more accurate interpretation of numbers and require lower numeracy, compared with alternative visualizations such as pie charts and bar charts~\cite{kay2016ish,galesic2009using}.}
\responseline{Users can inspect icon arrays at a glance to quickly understand how many OOD samples might exist in the test data.}
To validate if an optimal threshold is set, users can check the OOD score distribution (Fig.~\ref{fig:control_panel}~\circled{C}) across training and test data. Ideally, an optimal threshold should distinguish ID and OOD data as accurately as possible. {\tool} provides a pre-computed threshold to help non-experts efficiently decide the threshold.

\begin{figure*}[t]
    \centering
    \includegraphics[width=0.9\linewidth]{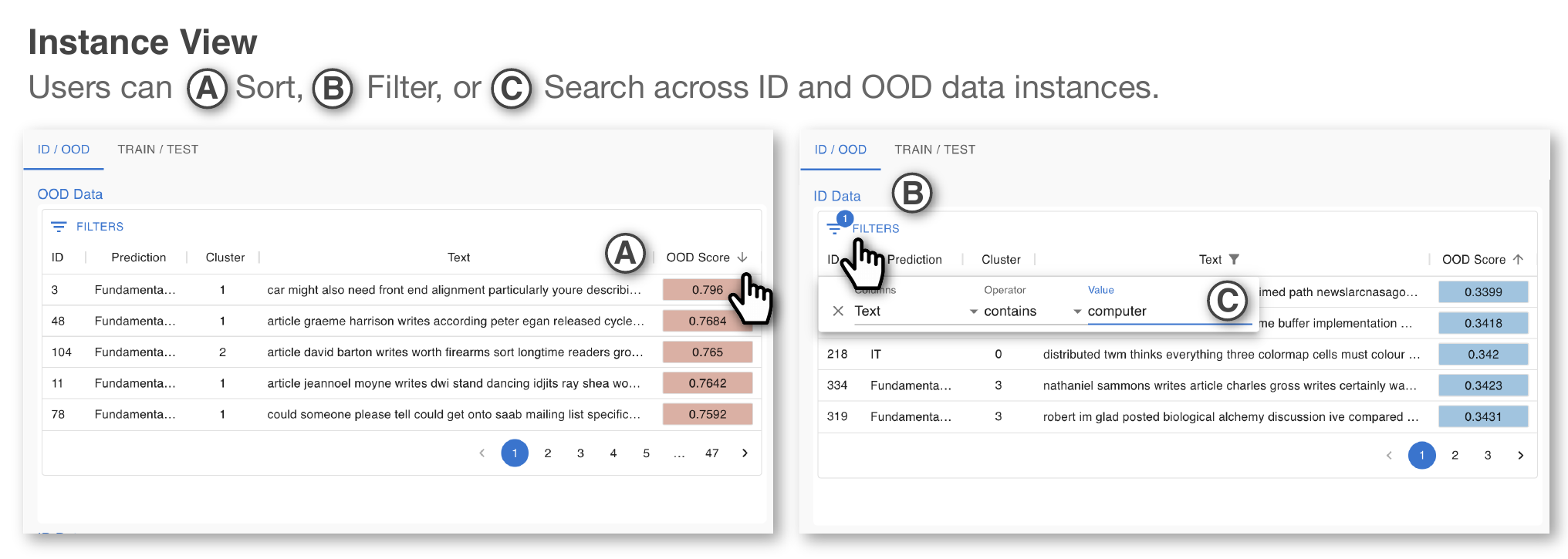}
    \caption{Interaction with the \textbf{\textit{Instance View}}.}
    \Description{This figure displays three components and two types of interactions in the Cluster View. Component A is a scatter plot showing all instances in the Cluster View with color-coded based on the different clusters. Component B is the nodes in the selected cluster, i.e., Cluster 1 in the figure. Component C is the keywords extracted from Cluster 1 instances. The first interaction is when users click on a specific cluster, i.e., Cluster 1 in the figure, the Cluster View only shows nodes in Cluster 1 and updates the keywords related to Cluster 1. The second interaction is when users hover over a specific node, i.e., a node with the prediction: Fundamental SCIENCE and OOD Score: 0.7631 in the figure, they can find the prediction label, cluster label, and OOD score of this node. }
    \label{fig:instance_view}
\end{figure*}

\vspace{1mm}
{\noindent \textbf{\textit{Instance View.}}} This view contains two separate scrollable data grids of the ID and OOD data (Fig.~\ref{fig:instance_view}). When users change the threshold of the OOD score, the instance view will update accordingly. The rows of each data grid are individual data instances, and the columns are: \textit{index}, \textit{model prediction result}, \textit{clustering result}, \textit{raw text}, and \textit{OOD score} of each data instance. By default, the data grids including \textit{OOD data} and \textit{ID data} are sorted in descending and ascending orders according to OOD scores, respectively. Users can also filter, search, or sort each data grid to explore a data instance. 

\subsection{OOD Text Categorization and Exploration}
\label{subsec:identify}

{\noindent \textbf{\textit{Text Clustering.}}} To help users efficiently explore topics of detected OOD samples, {\tool} uses a text clustering algorithm to categorize different types of texts. Algorithm~\ref{alg:cluster} depicts the clustering algorithm. Given an NLP model $M$, the algorithm first extracts hidden features for each text of new test data (Line 1-5). While the extracted features $\mathcal{F}$ are usually sparse and high-dimensional vectors, {\tool} applies \textit{PCA} (principal component analysis) \responseline{\cite{jolliffe2002principal}} to reduce their dimensions to $p$ (line 6). Then, {\tool} uses \textit{KMeans} clustering algorithm \responseline{\cite{lloyd1982least}} to cluster processed hidden features $\mathcal{F}_p$ (Line 7-9). To decide an optimal number of clusters $n_{opt}$, {\tool} leverages \textit{Silhouette} method~\cite{rousseeuw1987silhouettes} (Line 10). During implementation, we set the maximum number of clusters as $N_{max}=200$, and the PCA dimension as $p=128$. These numbers are decided empirically.


\vspace{1mm}
\sloppy {\noindent \textbf{\textit{Keywords Summarization.}}} After each cluster is determined, {\tool} summarizes a few keywords from an individual cluster to help users identify its potential topic(s). To achieve this, {\tool} first filters out ``stop words''~\cite{bird2009natural} from each data and then uses CountVectorizer algorithm~\cite{scikit-learn} to extract keywords. For each cluster, {\tool} displays top-10 frequent keywords as a word cloud. Through inspecting the word cloud, users can quickly understand what kind of text patterns the selected cluster might include.

\begin{algorithm}[t]
\caption{A semantic text clustering algorithm.}
\label{alg:cluster}
\KwIn{an NLP model $M$, new test data $\mathcal{X}$, maximum number of clusters $N_{max}$, PCA dimension $p$}
\KwOut{cluster label $\mathcal{Y}$}
$\mathcal{F}~\gets~\{\}$\;
\For{$x~$in$~\mathcal{X}$}{
$f~\gets~\mathsf{extract\_features}(M, x)$\;
$\mathcal{F}$.append$(f)$\;
}
$\mathcal{F}_p~\gets~\mathsf{PCA}(\mathcal{F}, p)$\;
\For{$n~\gets~1,\dots,N_{max}$}{
$\mathcal{Y}_n=\mathsf{KMeans(\mathcal{F}_p, n)}$\;
}
$n_{opt}~\gets$~$\displaystyle \max_n\mathsf{Silhouette}(\mathcal{Y}_n, \mathcal{F}_p)$\;
\KwRet{$\mathcal{Y}_{n_{opt}}$}\;
\end{algorithm}


\begin{figure*}[t]
    \centering
    \includegraphics[width=0.7\linewidth]{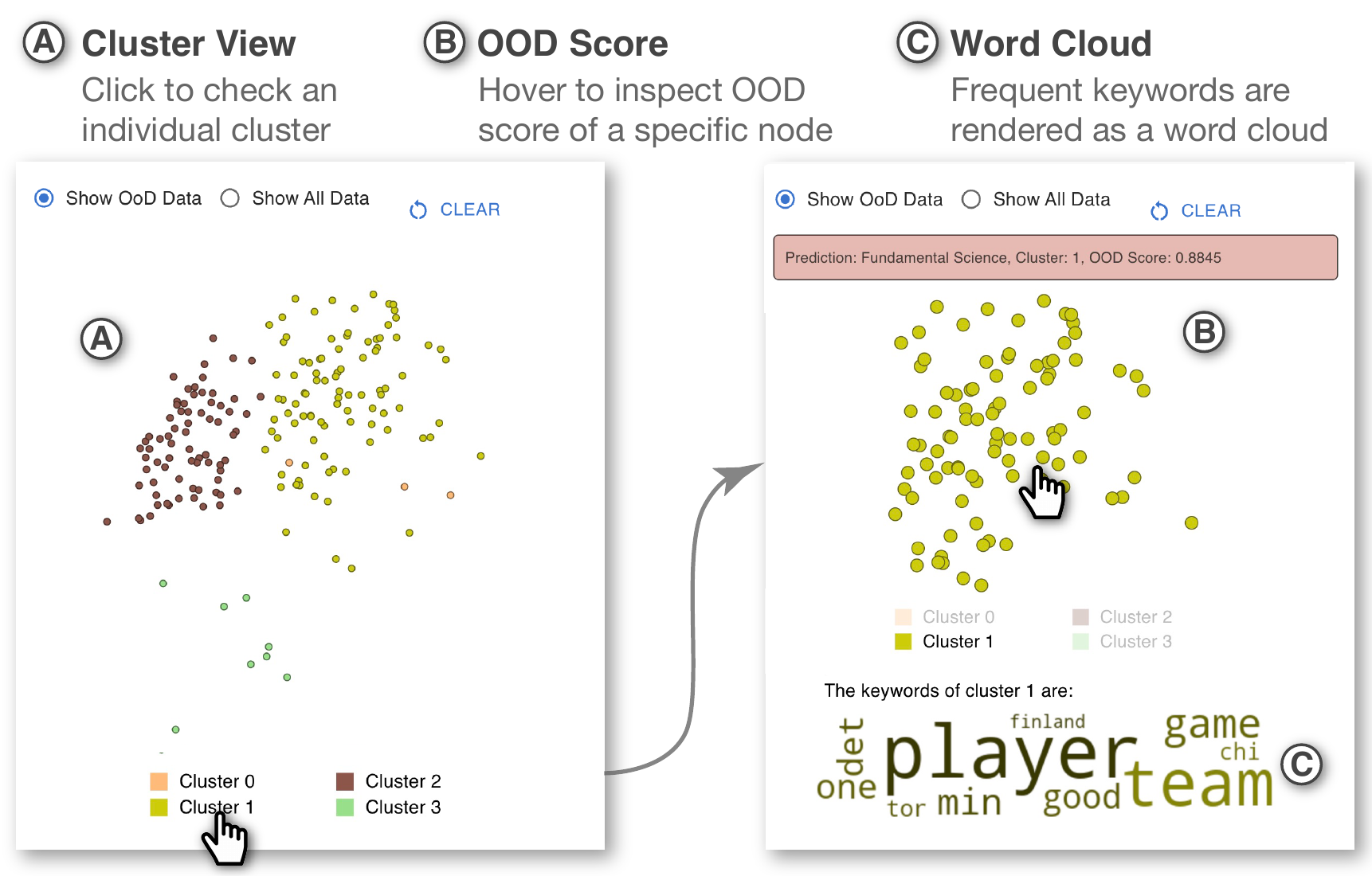}
    \caption{Users can inspect and interact with text clustering results in the \textbf{\textit{Cluster View}}.}
    \Description{This figure displays how the Highlighting View helps users quickly understand specific ID/OOD data instances. There are two components in the Highlighting View. One is the ID Samples stack, the other is the OOD Samples stack. The two components show all selected ID and OOD Samples with keywords highlighted. This Figure also contains the instruction on how to use the Highlighting View. There are three steps. First, click ID/OOD instances by pinning them at the top of the ID/OOD table. Then the Salient words are highlighted to ease the user’s reading efforts. Finally, users can compare highlighted keywords of ID/OOD instances side-by-side to find the difference between ID/OOD data.}
    \label{fig:cluster_view}
\end{figure*}

\vspace{1mm}
{\noindent \textbf{\textit{Cluster View.}}} {\tool} integrates the clustering results and summarized keywords in the cluster view (Fig.~\ref{fig:cluster_view}). The cluster view consists of a scatter plot and a word cloud. Each node in the scatter plot represents an individual data instance. \responseline{The position of each node is determined by the first three components of the hidden features of each text after PCA.} The color assigned to each node represents the cluster index. When users hover on a node, a tool-tip will pop up showing the prediction label and OOD score of the corresponding data instance (Fig.~\ref{fig:cluster_view}~\circled{B}). When users click on a node, the corresponding data instance will also be selected in the instance view. These features allow users to contextualize the clustering results with specific data instances and texts. Users can also focus on one cluster by clicking on the legend (Fig.~\ref{fig:cluster_view}~\circled{A}). Once a cluster is selected, {\tool} will update the word cloud (Fig.~\ref{fig:cluster_view}~\circled{C}) and filter out data excluded in the selected cluster in the instance view. 

\subsection{OOD Text Explanation}
\label{subsec:understand}

{\noindent \textbf{\textit{Salient Words Selection.}}} The previous sections introduce how users can efficiently inspect OOD issues and identify potential OOD types. By inspecting the cluster view, users might already have hypotheses about potential OOD types and their topics based on several data instances. {\tool} further supports digging into specific instance(s). To achieve this, {\tool} uses neuron activation analysis~\cite{hewitt2019designing} to select salient words in a text. By checking small groups of highlighted words, users can avoid reading a long paragraph of text in detail. We describe the algorithm of \textit{salient words selection} in Algorithm~\ref{alg:salient}. At a high level, {\tool} leverages ecco~\cite{alammar-2021-ecco} to extract and factorize neuron activation information. For a given text $x$, {\tool} first extract neuron activation values $\mathcal{A}$ by passing it through a large pre-trained language model $M$ (Line 2). Then {\tool} uses Non-negative matrix factorization to factorize the extracted activation values into $n$ components (Line 3). In this way, {\tool} can group similar words in a text into $n$ groups. To further reduce users' mental demands, we filter out groups containing stop words or special tokens (e.g., punctuation) (Line 4-14). Finally, for each group, {\tool} only highlights 10 words with the highest activation values. This helps preserve only the most important words in a group. In our implementation, \responseline{we use a pre-trained BERT~\cite{devlin2018bert} released on HuggingFace~\footnote{https://huggingface.co/bert-base-uncased} without any fine-tuning.} The number of factors $n$ is set to 10, and only top-10 salient words are highlighted.

\begin{algorithm}[t]
\caption{Salient words selection.}
\label{alg:salient}
\KwIn{A text input $x$, a pre-trained language model $M$, number of factors $n$, number of words in text $l$}
\KwOut{Highlighted keywords $\mathcal{S}$}
$\mathcal{S}~\gets~\emptyset$\;
$\mathcal{A}~\gets~\mathsf{get\_activation}(M, x)$\;
$\Tilde{\mathcal{S}}~\gets~\mathsf{NMF}(\mathcal{A}, n)$\;
\For{$i~\gets1,\dots,n$\;}{
    \For{$j~\gets1,\dots,l$\;}{
        \eIf{$x_j$ is stop word or special tokens}{
                $\mathcal{S}_i~\gets~\emptyset$\;
            }{
                $\mathcal{S}~\gets~\mathcal{S}~\bigcup~\Tilde{\mathcal{S}}_i$\;
            }
    }
}
\KwRet{$\mathcal{S}$}\;
\end{algorithm}

\begin{figure*}[t]
    \centering
    \includegraphics[width=0.9\linewidth]{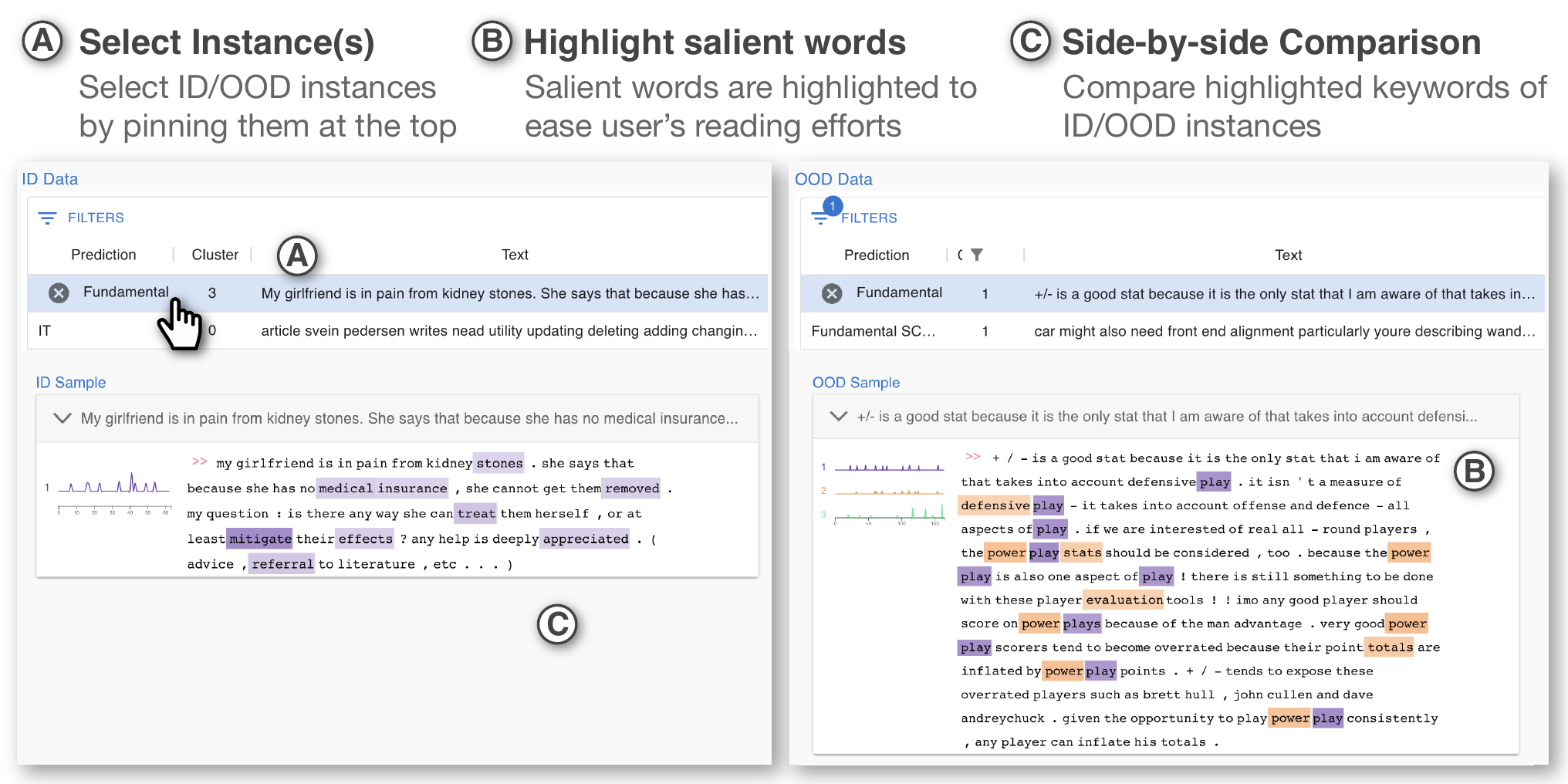}
    \caption{The \textbf{\textit{Highlighting View}} helps users quickly understand specific ID/OOD data instances.}
    \Description{This figure displays two components and three types of interactions in the Instance View. The two components are ID/OOD tables, which have 5 columns: ID, Prediction, Cluster, Text, and OOD Score. Interaction A is Sort, users can sort the OOD Score by ASC or DESC order. Interaction B and C is Filter, users can search the instances by words or filter by cluster label.}
    \label{fig:highlight_view}
\end{figure*}

\vspace{1mm}
{\noindent \textbf{\textit{Highlighting View.}}} In this view, {\tool} leverages the visualization of ecco~\cite{alammar-2021-ecco} to visualize highlighted salient words (Fig.~\ref{fig:highlight_view}). \responseline{The sparklines on the left of each text box visualize the positions of similar salient words in the text. Each sparkline represents a group of similar salient words. The color of a sparkline is the same as the color of the corresponding group of salient words. The x-axis is the index of a word in the text, and the y-axis indicates the saliency score of a word.} Users can hover on different lines to inspect different groups of salient words. For each group of salient words, a darker color indicates a higher activation value. While previous studies have shown that it is important to allow users to understand a concept through a contrastive way~\cite{miller2019explanation}, {\tool} allows users to pin multiple ID or OOD instances at the same time to contextualize the OOD topics. When users click a node in the \textit{cluster view} or an instance in the \textit{instance view}, this instance will be pinned at the top of the \textit{instance view}. At the same time, the salient words of each selected instance will be displayed in the \textit{highlighting view}.

\subsection{Implementation}
\label{subsec:implementation}
We implement and deploy {\tool} as a web application. The interface of {\tool} (Fig.~\ref{fig:interface}) is implemented with Material UI~\footnote{https://mui.com}. We use D3.js~\footnote{https://d3js.org} for visualizing scatter plots. All machine learning models were implemented with PyTorch and Scikit-learn and trained on one NVIDIA A6000 GPU. We deployed {\tool} on an AWS EC2 for ease of access during the user study.

\section{Usage Scenario}
\label{sec:scenario}

Suppose Alice is a model developer and she has trained an ML model to classify text documents into two different topics: \textit{IT (information technology)} and \textit{Fundamental Science}. Her model achieves 94\% accuracy on the training data. However, her model's performance significantly drops when Alice deploys it online. Alice suspects such performance degradation is caused by out-of-distribution (OOD) issues in the new data. Alice runs an OOD detection program on the new data and finds that 49\% of the new data (400 data instances) are detected as OOD data. Alice wants to understand the characteristics of these data, so she can strategically apply data augmentation techniques to improve the training data. However, she finds it time-consuming to manually check the 400 OOD samples and understand why they are categorized as OOD. Furthermore, since each text document is lengthy, Alice finds it hard to glance it over and quickly understand the gist of the document. 

Alice decides to give {\tool} a try. She first checks the \textit{Distribution View}, where two icon arrays (Fig.~\ref{fig:control_panel}~\circled{B}) showing the proportion of \textcolor[HTML]{188dc3}{in-distribution} and \textcolor[HTML]{c45a45}{out-of-distribution} samples in training and test data, respectively. She finds that compared with training data, a large proportion of test data are \textcolor[HTML]{c45a45}{out-of-distribution} samples. She also attends to the \textit{OOD score distribution} (Fig.~\ref{fig:control_panel}~\circled{C}), where the OOD score distribution of \textcolor[HTML]{69b3a2}{training} and \textcolor[HTML]{F7CB8A}{test} data are different. She confirms that OOD samples exist in her test data and could have caused her model's performance degradation.

Alice wonders what kinds of instances are detected as OOD samples by {\tool}. Therefore she turns to the \textit{Instance View} (Fig.~\ref{fig:interface} \circled{B}), where the OOD and ID data are displayed in two separate data grids. Alice finds that instances with high OOD scores are very suspicious. Since browsing each data instance will take too much time, Alice decides to switch to the \textit{Cluster View} (Fig.~\ref{fig:interface} \circled{C}) to get an overview first. This view contains a scatter plot where each data instance is rendered as a node, and similar data instances are clustered and colored with the same color. By default, the \textit{Cluster View} only shows OOD data. Alice finds it obvious that each cluster contains a different number of OOD nodes. Then she explores the exact number of OOD nodes in each cluster by hovering over the legends. There are many OOD nodes in Cluster 1 and 2, but much fewer in Cluster 0 and 3 (Fig.~\ref{fig:cluster_view}~\circled{A}). Alice thinks Cluster 1 and 2 may include two new OOD types.

Now Alice wants to take a deeper look at Cluster 1. By clicking the corresponding legend, the \textit{Cluster View} is updated and only displays OOD nodes of Cluster 1 (Fig.~\ref{fig:cluster_view}~\circled{B}). The \textit{Instance View} also filters out the data instances that are not in Cluster 1. The \textit{word cloud} shows that Cluster 1 has several frequent keywords, e.g., ``players'', ``game'', ``team'' (Fig.~\ref{fig:cluster_view}~\circled{C}). Alice realizes the topic of this cluster could be \textit{Sports}. This is an obvious sign of a new category in the OOD data, which is not initially included in the training data.

Alice clicks a node in Cluster 1. Then, the selected instance is pinned at the top of the \textit{Instance View}. While the selected sentence is very long and takes time to read, Alice decides to check the highlighted words in the \textit{Highlighting View} (Fig.~\ref{fig:highlight_view}~\circled{B}). These salient words help her focus on the essential information and ignore the unnecessary words in the sentence. Alice notices that there are a few words highlighted, e.g., ``defensive'', ``power play''. These keywords further confirm Alice's belief that a new category is \textit{Sports}.

To further validate her belief, Alice continues to check whether this category exists in the in-distribution data. She clicks on the first sentence in the ID table of the \textit{Instance View}, which is predicted as \textit{Fundamental Science} (Fig.~\ref{fig:highlight_view}~\circled{A}). The salient words highlighted in this sentence are: ``medical'', ``stones'', ``mitigate'', and ``treat'' (Fig.~\ref{fig:highlight_view}~\circled{C}). All these words are usually from \textit{Fundamental Science} articles. \responseline{Then, Alice clicks into several other OOD instances near the current OOD instance and confirms that \textit{Sports} is a new topic in the OOD data.}

\aptLtoX[graphic=no,type=html]{\begin{table*}[t]
    \centering
    \caption{Designed tasks for the user study.}
    \small
    \begin{tabular}{l l l l l l}
         \toprule
         \textbf{\#} & \textbf{Task} & \textbf{Description} & \textbf{ID data} & \textbf{OOD data} & \textbf{Distribution shift type} \\
         \midrule
         \multirow[t]{2}{*}{1} & \multirow[t]{2}{*}{Topic classification} & Predict topic of a paragraph & DBPedia & DBPedia & \multirow[t]{2}{*}{Semantic shift}\\
         & & of text from Wikipedia. & top-4~\cite{zhang2015character}$^\dagger$ & rest~\cite{zhang2015character}$^\dagger$ &\\
         \midrule
         \multirow[t]{2}{*}{2} & \multirow[t]{2}{*}{Sentiment analysis} & Predict sentiment of a & IMDB~\cite{maas2011learning} & Yelp~\cite{zhang2015character} & \multirow[t]{2}{*}{Background shift}\\
         & & review text. & & &\\
         \midrule
         \multirow[t]{2}{*}{3} & \multirow[t]{2}{*}{Emotion recognition} & Recognize the emotion & Emotion & Emotion & \multirow[t]{2}{*}{Semantic shift}\\
         & & from a given text. & Negative-2~\cite{saravia2018carer}$^\ddagger$ & rest~\cite{saravia2018carer}$^\ddagger$ &\\
         \midrule
         \multirow[t]{3}{*}{4} & \multirow[t]{3}{*}{Fake news detection} & Detect if a news article & PolitiFact~\cite{shu2020fakenewsnet} & COVID-19 & \multirow[t]{3}{*}{Background shift}\\
         & & is real or fake. & & Fake News~\cite{das2021heuristic} & \\
         & & & & GossipCop~\cite{shu2020fakenewsnet} & \\
         \bottomrule
        \multicolumn{6}{l}{$^\dagger$ DBPedia dataset has 14 classes. We denote DBPedia top-4 as a subset including the first 4 classes (Company, Educational Institution, Artist, and Athlete) according to class IDs, and DBPedia rest as a subset including the other 10 classes.}\\
        \multicolumn{6}{l}{$^\ddagger$ Emotion dataset has 6 classes. We denote Emotion Negative-2 as a subset including 2 negative Emotion classes (Sadness and Fear), and Emotion rest as a subset including the other 4 classes.}
    \end{tabular}
    \label{tab:task}
\end{table*}}{\begin{table*}[t]
    \centering
    \caption{Designed tasks for the user study.}
    \small
    \begin{tabular}{l l l l l l}
         \toprule
         \textbf{\#} & \textbf{Task} & \textbf{Description} & \textbf{ID data} & \textbf{OOD data} & \textbf{Distribution shift type} \\
         \midrule
         \multirow[t]{2}{*}{1} & \multirow[t]{2}{*}{Topic classification} & Predict topic of a paragraph & DBPedia & DBPedia & \multirow[t]{2}{*}{Semantic shift}\\
         & & of text from Wikipedia. & top-4~\cite{zhang2015character}$^\dagger$ & rest~\cite{zhang2015character}$^\dagger$ &\\
         \midrule
         \multirow[t]{2}{*}{2} & \multirow[t]{2}{*}{Sentiment analysis} & Predict sentiment of a & IMDB~\cite{maas2011learning} & Yelp~\cite{zhang2015character} & \multirow[t]{2}{*}{Background shift}\\
         & & review text. & & &\\
         \midrule
         \multirow[t]{2}{*}{3} & \multirow[t]{2}{*}{Emotion recognition} & Recognize the emotion & Emotion & Emotion & \multirow[t]{2}{*}{Semantic shift}\\
         & & from a given text. & Negative-2~\cite{saravia2018carer}$^\ddagger$ & rest~\cite{saravia2018carer}$^\ddagger$ &\\
         \midrule
         \multirow[t]{3}{*}{4} & \multirow[t]{3}{*}{Fake news detection} & Detect if a news article & PolitiFact~\cite{shu2020fakenewsnet} & COVID-19 & \multirow[t]{3}{*}{Background shift}\\
         & & is real or fake. & & Fake News~\cite{das2021heuristic} & \\
         & & & & GossipCop~\cite{shu2020fakenewsnet} & \\
         \bottomrule
        \multicolumn{6}{l}{$^\dagger$ DBPedia dataset has 14 classes. We denote DBPedia top-4 as a subset including the first 4 classes (Company, Educational Ins-}\\
        \multicolumn{6}{l}{titution, Artist, and Athlete) according to class IDs, and DBPedia rest as a subset including the other 10 classes.}\\
        \multicolumn{6}{l}{$^\ddagger$ Emotion dataset has 6 classes. We denote Emotion Negative-2 as a subset including 2 negative Emotion classes (Sadness and}\\
        \multicolumn{6}{l}{Fear), and Emotion rest as a subset including the other 4 classes.}
    \end{tabular}
    \label{tab:task}
\end{table*}}

\section{User Study}
\label{sec:user_study}
To evaluate the effectiveness and usability of {\tool}, we conducted a within-subjects user study with 24 programmers with various levels of machine learning expertise. To better understand the value of proposed features in {\tool}, we compared {\tool} with a variant of {\tool} as the baseline by disabling the \textit{Cluster View} and the \textit{Highlighting View}.

\subsection{Participants}

We recruited 24 participants through mailing lists of the ECE and CS departments at the University of Alberta~\footnote{This human-participated study is approved by the university’s research ethics office.}. All participants have basic knowledge about machine learning. 10 participants were Master students, 10 were Ph.D. students, 3 were professional developers, and 1 was a data scientist. Participants were asked to self-report their machine learning expertise. 12 participants had 2-5 years of experience, 1 had more than 5 years, and 11 only had 1 year. Regarding NLP experience, 5 participants had 2-5 years of experience, and 19 only had 1 year. 20 participants mentioned that they had heard about out-of-distribution or distribution shift problems before. All study sessions were conducted on Zoom. Both {\tool} and baseline were deployed as web applications, therefore participants were able to access our study sessions from their own PCs.

\subsection{Tasks}

We designed four tasks that cover different kinds of distribution shifts in the NLP domain. Table~\ref{tab:task} shows the details of each user study task. When designing tasks, we follow these requirements: (1) the tasks should be representative ones in the NLP domain and (2) the tasks should cover two different types of distribution shift. To achieve these goals, we collected four tasks from prior work about OOD detection in NLP and well-known public benchmarks for NLP models. For each task, we adopt a BERT model~\cite{devlin2018bert} as the backbone and fine-tune its performance on the ID data. \responseline{For a fair comparison, the baseline tool and {\tool} use the same pre-computed OOD threshold on each task.} More details such as models' training settings and example interfaces for each task can be found in Appendix~\ref{appendix:interface}.

\subsection{Protocol}

Each user study session took about 60 minutes. At the beginning of each session, we asked participants for their consent to record. Participants were assigned two tasks about identifying OOD issues, one to be completed with {\tool} and the other to be the baseline tool. To mitigate the learning effect, both task assignment order and tool assignment order were counterbalanced across participants. In total, 6 participants experienced each task. Participants were asked to watch a 5-min tutorial video of the assigned tool before starting each task, followed by a 5-min practice period to familiarize themselves with the tool. Then, participants were given 20 minutes to use the assigned tool to inspect and identify OOD issues within the given model and dataset. In particular, participants were asked to answer/report: 

\begin{enumerate}[leftmargin=*]
    \item What kind of data distribution shift does it belong to?
    \item How many different types of OOD data did you find?
    \item For each different type of OOD, please explain why you think it is OOD and list the indexes of the OOD instance that belong to this type.
\end{enumerate}

After completing each task, participants filled out a post-task survey to give feedback about what they liked or disliked. Participants were also asked to answer five NASA Task Load Index (TLX) questions~\cite{hart1988development} as a part of the post-task survey. After completing both two tasks, participants filled out a final survey, where they directly compared two assigned tools. At the end of the study session, each participant received a \$25 Amazon gift card as compensation for their time.

\section{Results}
\label{sec:results}

\begin{table*}[t]
    \caption{User performance in four different tasks.}
    \small
    \centering
    \begin{tabular}{lcccccccccccc}
         \toprule
         & & \multicolumn{2}{c}{\textbf{Task 1}} && \multicolumn{2}{c}{\textbf{Task 2}} && \multicolumn{2}{c}{\textbf{Task 3}} && \multicolumn{2}{c}{\textbf{Task 4}}\\
         & & \multicolumn{2}{c}{Semantic Shift} && \multicolumn{2}{c}{Background Shift} && \multicolumn{2}{c}{Semantic Shift} && \multicolumn{2}{c}{Background Shift}\\
         \cline{3-4} \cline{6-7} \cline{9-10} \cline{12-13}
         & & Baseline & {\tool} && Baseline & {\tool} && Baseline & {\tool} && Baseline & {\tool}\\
         \midrule
         \multicolumn{2}{l}{\# of participants correctly} & \multirow{2}{*}{5} & \multirow{2}{*}{6} & & \multirow{2}{*}{5} & \multirow{2}{*}{6} & & \multirow{2}{*}{6} & \multirow{2}{*}{6} & & \multirow{2}{*}{6} & \multirow{2}{*}{6}\\
         \multicolumn{2}{l}{identified shift type} & & & & & & & & & & &\\
         \hline
         \# of correct & Min & 1/10 & 4/10 && 1/2 & 1/2 && 0/4 & 3/4 && 0/2 & 2/2\\
         OOD Types& \cellcolor{lightgray}Med & \cellcolor{lightgray}2/10 & \cellcolor{lightgray}7.5/10 &\cellcolor{lightgray}& \cellcolor{lightgray}1/2 & \cellcolor{lightgray}2/2 &\cellcolor{lightgray}& \cellcolor{lightgray}1/4 & \cellcolor{lightgray}3.5/4 &\cellcolor{lightgray}& \cellcolor{lightgray}0.5/2 & \cellcolor{lightgray}2/2\\
         found per& Max & 4/10 & 8/10 && 1/2 & 2/2 && 2/4 & 4/4 && 1/2 & 2/2\\
         participant& \cellcolor{lightgray}Mean & \cellcolor{lightgray}2.3/10 & \cellcolor{lightgray}7/10 &\cellcolor{lightgray}& \cellcolor{lightgray}1/2 & \cellcolor{lightgray}1.7/2 &\cellcolor{lightgray}& \cellcolor{lightgray}1.2/4 & \cellcolor{lightgray}3.5/4 &\cellcolor{lightgray}& \cellcolor{lightgray}0.5/2 & \cellcolor{lightgray}2/2\\
         \cellcolor{tab_purple}&\cellcolor{tab_purple}&\multicolumn{2}{c}{\cellcolor{tab_purple}$\Delta=4.7$} &\cellcolor{tab_purple}& \multicolumn{2}{c}{\cellcolor{tab_purple}$\Delta=0.7$} &\cellcolor{tab_purple}& \multicolumn{2}{c}{\cellcolor{tab_purple}$\Delta=2.3$} &\cellcolor{tab_purple}& \multicolumn{2}{c}{\cellcolor{tab_purple}$\Delta=1.5$}\\
         \multicolumn{2}{c}{\cellcolor{tab_purple}\multirow{-2}{*}{Welch's $t$-test}}& \multicolumn{2}{c}{\cellcolor{tab_purple}$t=6.14,~p<0.001$} &\cellcolor{tab_purple}& \multicolumn{2}{c}{\cellcolor{tab_purple}$t=3.16,~p=0.010$} &\cellcolor{tab_purple}& \multicolumn{2}{c}{\cellcolor{tab_purple}$t=6.14,p<0.001$} &\cellcolor{tab_purple}& \multicolumn{2}{c}{\cellcolor{tab_purple}$t=6.71,p=0.001$}\\
         \bottomrule
    \end{tabular}
    \label{tab:task_performance}
\end{table*}

In this section, we report and analyze the difference in participants' performance when using {\tool} and the baseline tool. We denote the participant as P\# in the following.

\subsection{User Performance}

Table~\ref{tab:task_performance} shows participants' performance on four tasks of identifying OOD issues. We found that all 24 participants using {\tool} correctly identified the type of data distribution shift (i.e., background vs.~semantic shift) in the assigned model, while 2 out of 24 participants using the baseline method failed.

To further assess participants' performance, we manually inspected participants' answers to check their correctness. A correct OOD type should (1) include the word(s) that are representative of a group of data instances, and (2) be significantly different from ID data. Overall, we found that participants using {\tool} were able to find more types of OOD on all four different tasks compared with participants using the baseline tool. Regarding Task 1 and Task 3 (semantic shift), the average number of OOD types found by each participant using {\tool} is 7 and 3.5 respectively. By contrast, the average number is 2.3 and 1.2 respectively when using the baseline tool. The Welch's $t$-test suggests that the performance differences are significant in both cases ($p\text{-value}<0.001$). Regarding Task 2 and Task 4 (background shift), most participants using the baseline tool were only able to find 1 type of OOD (mean: 1 and 0.5 respectively). By contrast, participants using {\tool} were able to find 1.7 and 2 types of OOD on average in these two tasks respectively. The Welch's $t$-test suggests that the performance differences are significant ($p\text{-value}=0.010$ and $p\text{-value}=0.001$). 

We analyzed the post-task survey responses and the recordings to understand why participants using {\tool} performed better. We found that {\tool} users' success mainly come from the \textit{Cluster View} and the \textit{Highlighting View}.
First, the cluster view significantly sped up the process of finding OOD types. 23 out of 24 participants had heavily utilized the clustering view to explore OOD data. By contrast, participants using the baseline tool had to inspect OOD instances one by one. P16 wrote, ``\textit{[when using the baseline tool], it is tedious to go through all the data point one by one especially when there are a lot of them.}'' In the post-task survey, 22 out of 24 participants also agreed that the clustering results were helpful. P17 said, ``\textit{by going through clusters, I can find trends faster than by going through individual data points.}'' Besides, the summarized keywords of each cluster were also found helpful. Based on the recordings, 19 out of 24 participants started their exploration from these keywords. By contrast, participants using the baseline tool usually started their exploration by randomly picking an instance. P20 said, ``\textit{[when using DeepLens,] I can use the keywords extracted in the cluster and put that in the filter to find more OOD instances of the same type.}'' P21 commented, ``\textit{In my using experience, [DeepLens] helps me a lot in quickly summarizing background shift keywords.}

In addition, the highlighting view in {\tool} helped participants avoid incorrect OOD types. In our user study, the median number of incorrect OOD types found per participant using {\tool} is 0, while the corresponding number of participants using the baseline tool is 1. The mean difference of incorrect OOD types is 0.52 vs. 1.64 (Welch's $t$-test: $p=0.009$). One specific reason is that when using {\tool}, participants were able to compare the OOD data with the ID data to confirm a new OOD type. In the post-task survey, 17 out of 24 participants marked the comparison of ID and OOD data as helpful. 
Furthermore, when comparing the ID and OOD data, the highlighted keywords also helped participants avoid misunderstanding a long text. In the post-task survey, 18 out of 24 participants agreed that seeing the highlighted keywords was helpful. P9 said, ``\textit{when verifying my hypothesis about whether a certain sentence belongs to OOD data, DeepLens is helpful because it shows several highlighted keywords and reduces my time consumption.}'' By contrast, P11 commented, ``\textit{It is not easy to read the whole text [when using the baseline tool].}''

\responseline{We have also further analyzed the impact of the OOD threshold adjustment feature on user performance. We found that only two users (P6 and P14) had tried to adjust the threshold, and they eventually reset it to the pre-computed one. A plausible explanation is that the default threshold has already provided a good starting point for users to investigate OOD issues. Thus, we believe the OOD threshold in the distribution view may have little impact on user performance. Since this threshold adjustment feature is present in both conditions, the better user performance of DeepLens comes directly from the cluster and highlighting views.}

\subsection{User Confidence and Cognitive Overhead}

\begin{figure}[t]
    \centering
    \includegraphics[width=0.9\linewidth]{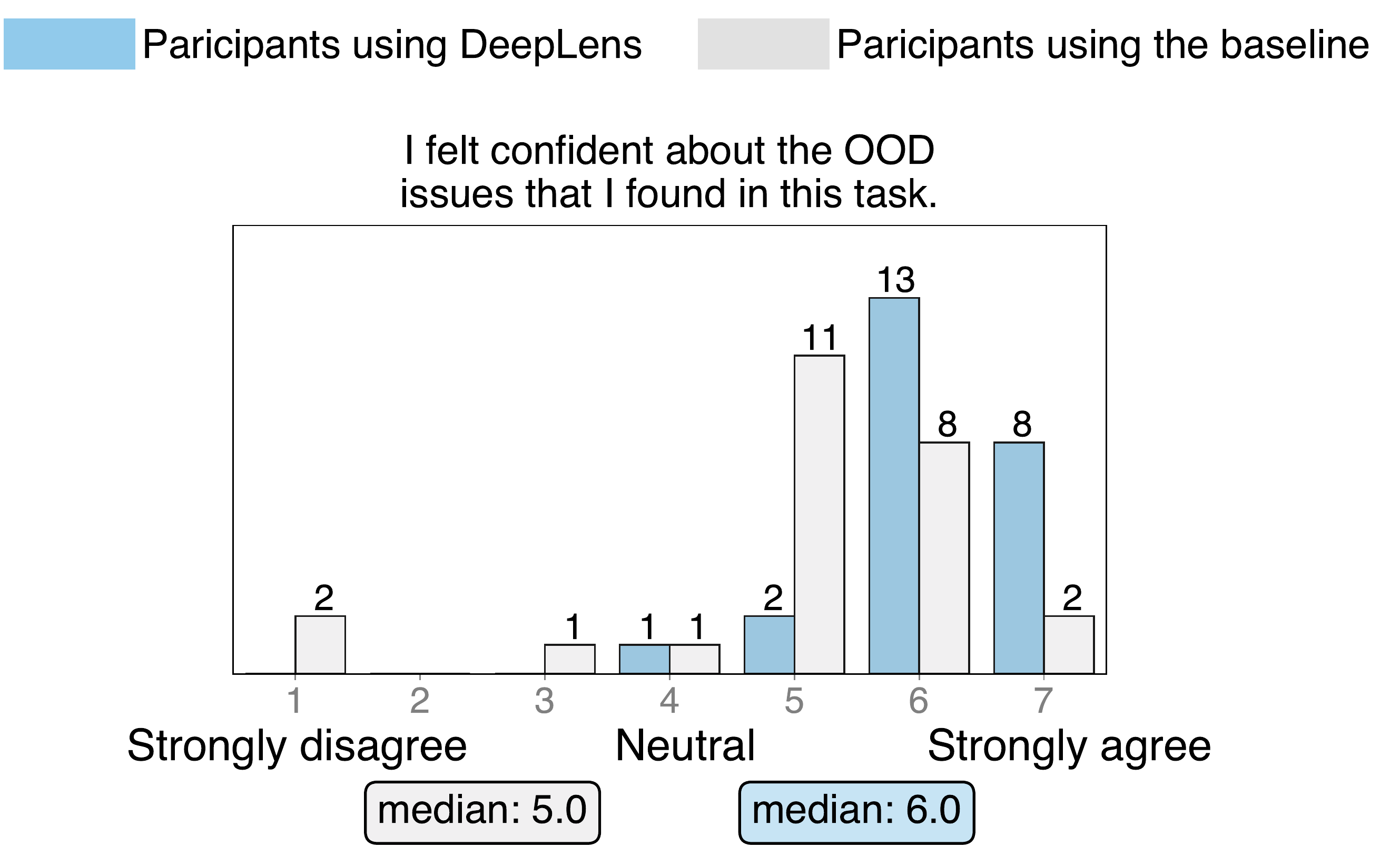}
    \caption{Users' self-ratings about their task performance.}
    \label{fig:self_rating}
    \Description{This figure shows the participates rating from two points about their task performance after using DeepLens and the Baseline tool. The first point is that I have a better understanding of the OOD issues after using the given tool. The second point is that I felt confident about the OOD issues taht I found in this task. DeepLens’s median rating is 6 on both points and the Baseline tool’s rating is 5 on both points. This figure shows that the participants using DeepLens have better understanding of of the OOD issues and felt confident about the OOD issues that they found in this task.}
\end{figure}

In the post-task survey, participants self-reported their confidence about OOD issues they identified with help of the assigned tool in two different 7-point Likert scale questions. Figure~\ref{fig:self_rating} shows participants' assessments when using {\tool} and the baseline tool. 
We found that participants using {\tool} were more confident about the OOD issues they found, where the median confidence ratings are 6 vs. 5. The mean difference is 1.12 (6.17 vs. 5.04), which is statistically significant (Welch's $t$-test: $p\text{-value}=0.002$). This confidence improvement was largely attributed to {\tool}'s \textbf{\textit{Cluster View}}. P5 commented, ``\textit{The automatic clustering function works quite well, and the keyword summary is quite useful to have an overview.}'' P16 said, ``\textit{By clustering the data, DeepLens makes keywords in the word cloud better indicators when identifying OOD issues.}

\begin{figure}[t]
    \centering
    \includegraphics[width=0.95\linewidth]{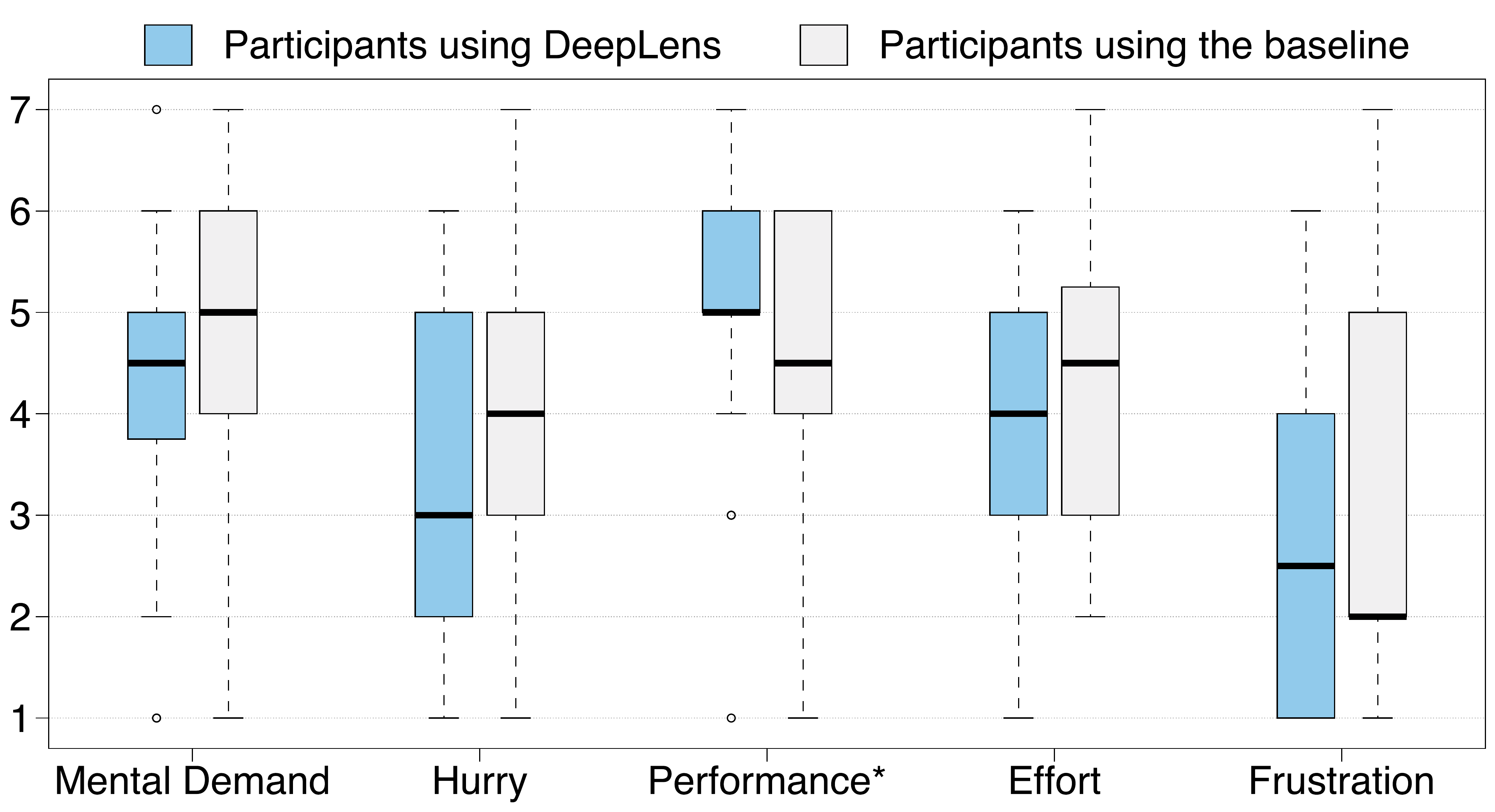}
    \caption{NASA Task Load Index Ratings. Entries with a star (*) mean the mean difference is statistically significant.}
    \Description{This figure shows the cognitive load of participants when using DeepLens versus the Baseline tool. The cognitive load is measured in five different dimensions including mental demand, how hurry participants feel during the study, participants’ self-assessment of their performance, participants’ self-assessment of the effort needed to complete the task, and how frustrated participants feel. There was no significant difference when using DeepLens and the Baseline tool though DeepLens has more features and renders more information.
}
    \label{fig:nasa_tlx}
\end{figure}

Figure~\ref{fig:nasa_tlx} shows participants' ratings on the five cognitive factors of the NASA TLX questionnaire. Though {\tool} has more features and renders more information, we find that there was no significant difference when using {\tool} and the baseline tool in terms of mental demand, hurry, effort, and frustration (Welch's $t$-test: $p=0.22$, $p=0.24$, $p=0.26$, $p=0.75$). However, participants using {\tool} felt they have better performance compared with participants using the baseline tool (mean difference: $0.88$, Welch's $t$-test: $p=0.049$). This indicates that {\tool} is more effective and useful when helping users inspect and identify OOD issues in an ML model compared with the baseline tool.

\subsection{User Ratings of Individual Features}

\begin{figure*}[t]
    \centering
    \includegraphics[width=\linewidth]{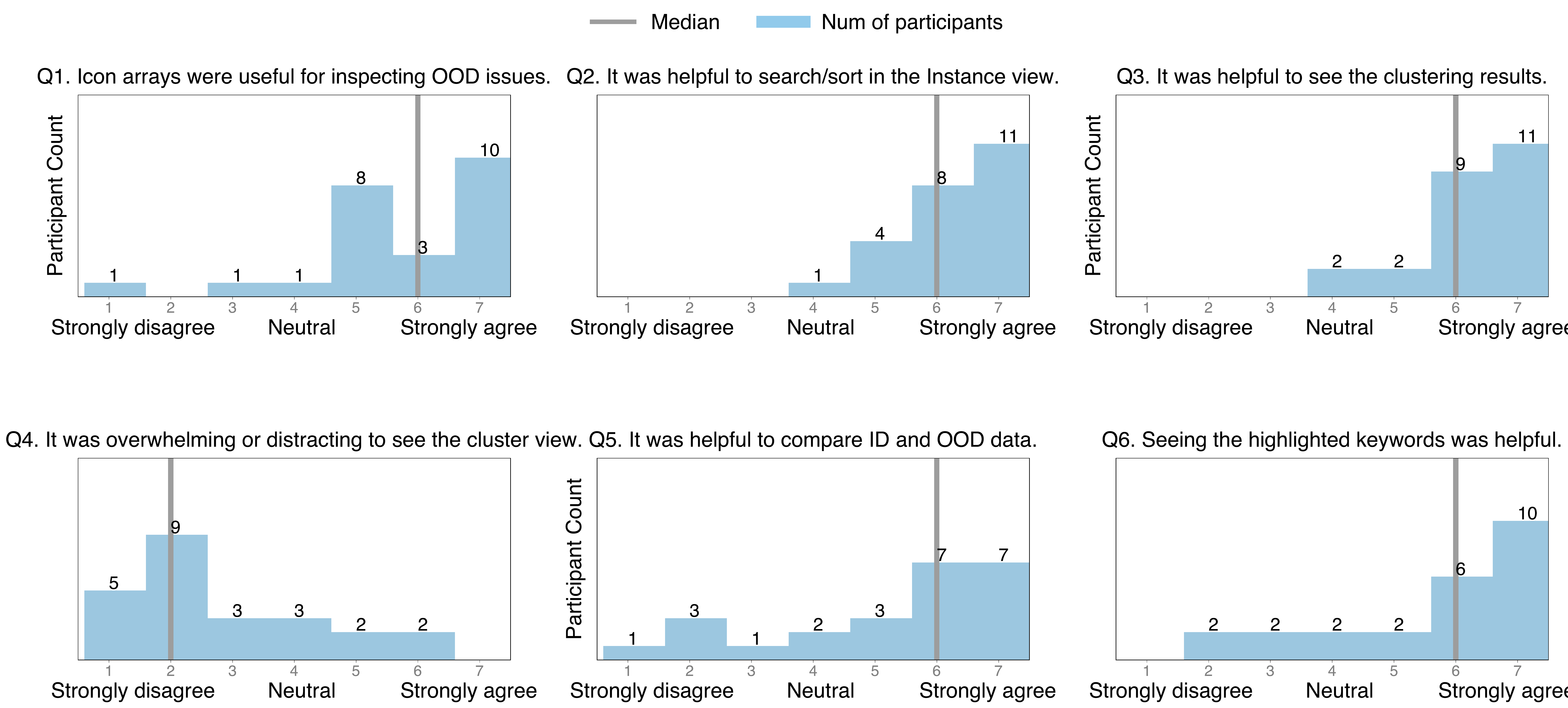}
    \caption{Users' ratings on individual features of {\tool}.}
    \Description{This figure shows the participants’ ratings about tool features of DeepLens on a 7-point Likert scale (1 means strongly disagree and 7 means strongly agree). Figure 9 shows that all of our tool features are highly appreciated by the participants. The median rating for “Icon arrays were useful for inspecting OOD issues” is 6. The median rating for “It was helpful to search/sort in the Instance view” is 6. The median rating for “It was helpful to see the clustering results” is 6. The median rating for “It was helpful to compare ID and OOD data is 6”. The median rating for “Seeing the highlighted keywords was helpful” is 6. }
    \label{fig:feature_ratings}
\end{figure*}

\sloppy In the post-task survey, participants rated the key features of {\tool}. Among 24 participants, 23 participants indicated that they would like to use {\tool} when solving OOD problems in their own ML models, while 1 participant stayed neutral. The median rating is 6 on a 7-point Likert scale (1---I don't want to use it at all, 7---I will definitely use it). As shown in Fig.~\ref{fig:feature_ratings}, participants felt {\tool}'s interface and interactive features intuitive and helpful. The \textbf{\textit{Cluster View}} is most appreciated by participants. 22 out of 24 participants agreed that ``\textit{it was helpful to see the clustering results.}'' The median rating of the cluster view is 6. P17 commented, ``\textit{I like that it [DeepLens] had clustering, for instance, it was super quick to find "covid" and "entertainment" OOD categories using this feature.}'' P20 commented, ``\textit{DeepLens has the cluster and keyword visualization which can help me identify a type of OOD quickly.}'' Besides, 18 out of 24 participants agreed that ``\textit{seeing highlighted keywords was helpful.}'' The median rating of the \textbf{\textit{Highlighting View}} is 6. P24 commented in the post-task survey, ``\textit{[DeepLens] brings me less reading and easy to focus on the details}'' 17 out of 24 participants also found comparing ID and OOD data helpful (median rating: 6). 

\subsection{User Preference and Feedback}
\label{subsec:feedback}

\begin{figure*}[t]
    \centering
    \includegraphics[width=0.8\linewidth]{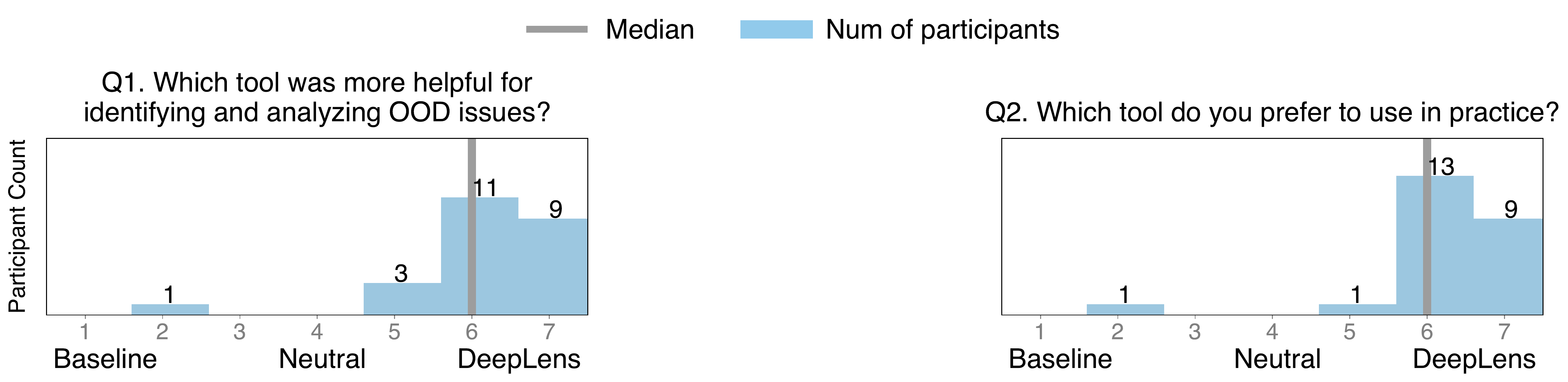}
    \caption{Users' preference on tool choice for OOD detection.}
    \Description{This figure shows users’ preference on tool choice for OOD detection. There are two questions in this figure. The first question is which tool was more helpful for identifying and analyzing OOD issues. The second question is which tool do you prefer to use in parctice. Both questions on a 7-point Likert scale (1 means strongly prefer to choose the Baseline tool and 7 means strongly prefer to choose DeepLens). The median of users’ rating on both questions is 6. And  23 out of 24 participants reported that DeepLens was more helpful and more preferred.
}
    \label{fig:user_preference}
\end{figure*}

In the final survey, participants self-reported their preference between {\tool} and the baseline tool (Figure~\ref{fig:user_preference}). 23 out of 24 participants reported that {\tool} was more helpful (median rating: 6) and they preferred to use it in practice (median rating: 6). We coded participants' responses to this question and identified 2 different themes. First, 17 participants mentioned that the \textbf{\textit{Cluster View}} makes it easier when identifying and analyzing OOD issues. P23 commented, ``\textit{keywords [in the cluster view] are useful when facing a large dataset.}'' P13 said, ``\textit{[the cluster view] gives a visual representation of the data, which makes it easier to identify the OOD data.}'' Second, 5 participants credited their success to the \textbf{\textit{Highlighting View}}. P9 said, ``\textit{highlighted keywords reduce my time consumption and make it easy to tell whether a data instance is OOD.}'' 

Participants also pointed out some limitations in the current form of {\tool}. 2 participants commented that it would be better if the keywords summarization in {\tool} could be improved. P21 said, ``\textit{there are some meaning-less high-frequency words which might disturb.}'' 4 participants suggested improving the usability of filters in the instance view, \eg, by allowing users to add multiple filter conditions at the same time. 1 participant mentioned that a semantic word search function (i.e., matching semantically similar words instead of identical ones) would assist their exploration process.

\section{Discussion}
\label{sec:discussions}

\subsection{Design Implications}


The user study results suggest that, with the help of {\tool}, users are able to find more types of OOD data with more confidence compared with using the baseline tool. Though addressing OOD issues is an urgent topic for deploying safe and reliable AI services~\cite{ren2019likelihood}, most efforts have been devoted to improving the algorithm accuracy of OOD detection. Our work indicates that only detecting OOD samples is not sufficient for improving model developers' productivity, especially when the dataset is large and the types of OOD are diverse. 
It is equally important to facilitate developers to understand and explore different types of OOD data in large text corpora. Once developers have gained deep insights of the OOD data in their datasets, they can further make strategic decisions to improve the model, e.g., data augmentation or selection for model retraining.

During the continuous delivery of machine learning models, the number of newly collected data can be massive. To reduce the cognitive effort of exploring different types of OOD instances in the new data, it is essential to summarize a small number of potential OOD types for developers. {\tool} addresses this by leveraging a text clustering algorithm. 
Furthermore, the interactive cluster exploration support in {\tool} preserves the user's control over verifying each type of OOD data. This is aligned with one of the human-AI interaction guidelines---\textit{providing several suggestions instead of fully automating the process}~\cite{amershi2019guidelines}. Based on the user study results, we find that such a semi-automated process of exploring OOD data improves participants' performance and confidence.

While clustering results do not directly tell users what exactly an OOD type is, the summarized keywords from each cluster serve as the starting point for exploration. In the final-study survey of our user study, 7 out of 23 participants who preferred to use {\tool} in practice explicitly mentioned how summarized keywords had assisted them. Previous work~\cite{chen2020oodanalyzer} has shown that one important requirement when designing an interactive system for OOD detection in image data is \textit{examining OOD samples in the context of normal samples.} {\tool} supports this by allowing users to compare ID text and OOD text side by side. Furthermore, compared with image data, text documents are less glanceable. Therefore, {\tool} highlights the salient words in each text document to help developers quickly grasp the gist of each document.

\subsection{Target User Groups and Use Cases}


{\tool} is designed for users who know ML but are not familiar with OOD issues. In our user study, most participants have heard about OOD issues but have not worked on OOD issues before. 4 out of 24 participants even reported that they had never heard about OOD before. Our user study results suggest that these participants performed better when using {\tool} compared with using the baseline tool (mean number of OOD types found per participant: 3.5 vs. 1.3, Welch's $t$-test: $p<0.0001$). Furthermore, they also felt more confident with the OOD issues they identified (median rating of confidence: 6 vs. 5). 
While experts may be more likely to identify OOD types by reading raw text data, they still appreciated {\tool} since it automates some of their work. For instance, the clustering in {\tool} automates the process of categorizing similar texts for them. P17 wrote, ``\textit{by going through these clusters, I can find trends faster than by going through individual data points.}'' P5 commented, ``\textit{DeepLens automated some of the manual work, and I found that my productivity is improved. I can get more work done within the same amount of time.}''

{\tool} is specifically designed for debugging OOD issues for NLP models. Therefore, our findings and design implications cannot be generalized to other kinds of ML issues, such as gradient vanishing. 
In addition, {\tool} can also be deployed as an online tool to continuously monitor potential data distribution shifts for deployed models. 

\subsection{Limitations and Future Work}

In addition to the limitations and suggestions pointed out by our user study participants (Sec.~\ref{subsec:feedback}), there are several other limitations to our user study design and system.

\vspace{1mm}
{\noindent \textbf{\textit{User Study Baseline}}.} In our current form of user study, a variant of {\tool} was created as the baseline method by disabling the cluster view and the highlighting view. However, this cannot distinguish the contribution of individual features to user performance improvement. One can consider creating more variants of {\tool} by disabling individual features as the comparison baselines. One can also consider instrumenting the tool and measuring the utility rate of each feature during user study sessions.

\vspace{1mm}
\sloppy {\noindent \textbf{\textit{Limited NLP Tasks}}.} Our user study cannot confirm whether {\tool} works for all types of NLP tasks. To comprehensively evaluate the usefulness of {\tool}, one can consider using {\tool} to identify OOD issues in more diverse NLP tasks, e.g., question answering and natural language inference.

\vspace{1mm}
\sloppy {\noindent \textbf{\textit{Accuracy of OOD Detection}}.} Currently, {\tool} leverages \textit{MSP}~\cite{arora2021types} as the OOD detection algorithm. Though \textit{MSP} has been proven effective in several NLP tasks~\cite{arora2021types}, it may not always be applicable to other kinds of NLP tasks or models. 
Since the design of {\tool} is not limited to a specific type of OOD detection algorithm, one future direction could be to integrate more OOD detection algorithms to {\tool} and allow users to switch between different algorithms.

\vspace{1mm}
{\responseref{}
{\noindent \textbf{\textit{Scalability Issue}}.} Based on our user study results, {\tool} can handle 1,000-4,500 data points. However, once the data is scaled up (\eg, millions of data points), nodes in the cluster view may overlap with each other. To address this issue, one can leverage more advanced visualization techniques such as Bubble Treemaps \cite{gortler2018visual} to visualize clustering results hierarchically.

\vspace{1mm}
{\noindent \textbf{\textit{Alternative Algorithms and Design}}.} Our cluster view can be further improved by using more advanced dimension reduction and clustering algorithms. In the current version of {\tool}, we choose PCA for dimension reduction and K-Means for text clustering since they are classical and common choices. However, more advanced dimension reduction methods, e.g., t-SNE~\cite{hinton2002stochastic}, Isomap~\cite{tenenbaum2000global} could potentially lead to better dimension reduction results. Besides, our text clustering can also be improved with methods that are specialized for topic modeling, e.g. ConceptScope~\cite{zhang2021conceptscope} and TopicNets~\cite{gretarsson2012topicnets}. \camerareadyrevision{Finally, our highlighting view can potentially be improved by replacing neuron activation analysis with other interactive tools for selecting and visualizing salient words in text data, e.g., exBERT~\cite{hoover2019exbert}.}

}

\section{Conclusion}
\label{sec:conclusion}

In this paper, we present a novel interactive system, {\tool}, to help ML developers detect, explore, and understand potential OOD (out-of-distribution) issues in NLP models. {\tool} leverages a text clustering algorithm to help users efficiently identify and explore potential types of OOD in large-scale text data. Furthermore, {\tool} integrates a neuron activation analysis-based algorithm to highlight salient words in an individual data instance to help users quickly understand a text without reading it in detail. We implemented {\tool} as a web application and conducted a within-subjects user study with 24 ML developers on four different NLP tasks. The results show that with the help of {\tool}, developers were able to have a better understanding of OOD issues in ML models and identify more types of OOD data confidently compared with using the baseline tool. In the end, we discuss the design implications from {\tool} and propose several promising future directions.

\begin{acks}
We would like to thank all anonymous participants in the user study and anonymous reviewers for their valuable feedback. This work was supported in part by Amii RAP program, Canada CIFAR AI Chairs Program, the Natural Sciences and Engineering Research Council of Canada (NSERC No.RGPIN-2021-02549, No.RGPAS-2021-00034, No.DGECR-2021-00019), as well as JSPS KAKENHI Grant No.JP20H04168, JST-Mirai Program Grant No.JPMJMI20B8.
\end{acks}







\bibliographystyle{ACM-Reference-Format}
\bibliography{reference}










\setcounter{figure}{0}
\def\thefigure{\Alph{section}\arabic{figure}}

\balance

\appendix

\section{Tasks for User Study}
\label{appendix:interface}

\subsection{NLP Task 1: Topic Classification}

DBPedia dataset extracts structured content from the information created in the Wikipedia project. In this task, we use DBPedia-14, collected by picking 14 non-overlapping topics from Wikipedia in 2014. We use examples from the first 4 classes as ID data and the rest as OOD data. To simulate real-world data with semantic shift, we sample 1000 instances from the test splits of 14 topics.  

\noindent\textbf{In-distribution Data}
ID data contains \responseline{300 samples and} 4 topics: Company, Educational Institution, Artist, and Athlete.

\noindent\textbf{Out-of-distribution Data}
OOD data contains \responseline{700 samples and} 10 topics: Office Holder, Mean of Transportation, Building, Natural Place, Village, Animal, Plant, Album, Film, and Written Work.

\noindent\textbf{Model}
We fine-tune the BERT model for 1 epoch on the training split of ID data with a learning rate of $5\times10^{-5}$ and a batch size of 16. It achieves 98.5\% accuracy on the validation splits of ID data and 28\% accuracy on the \responseline{test} data.

\begin{figure*}[t]
  \centering
  \includegraphics[width=0.85\linewidth]{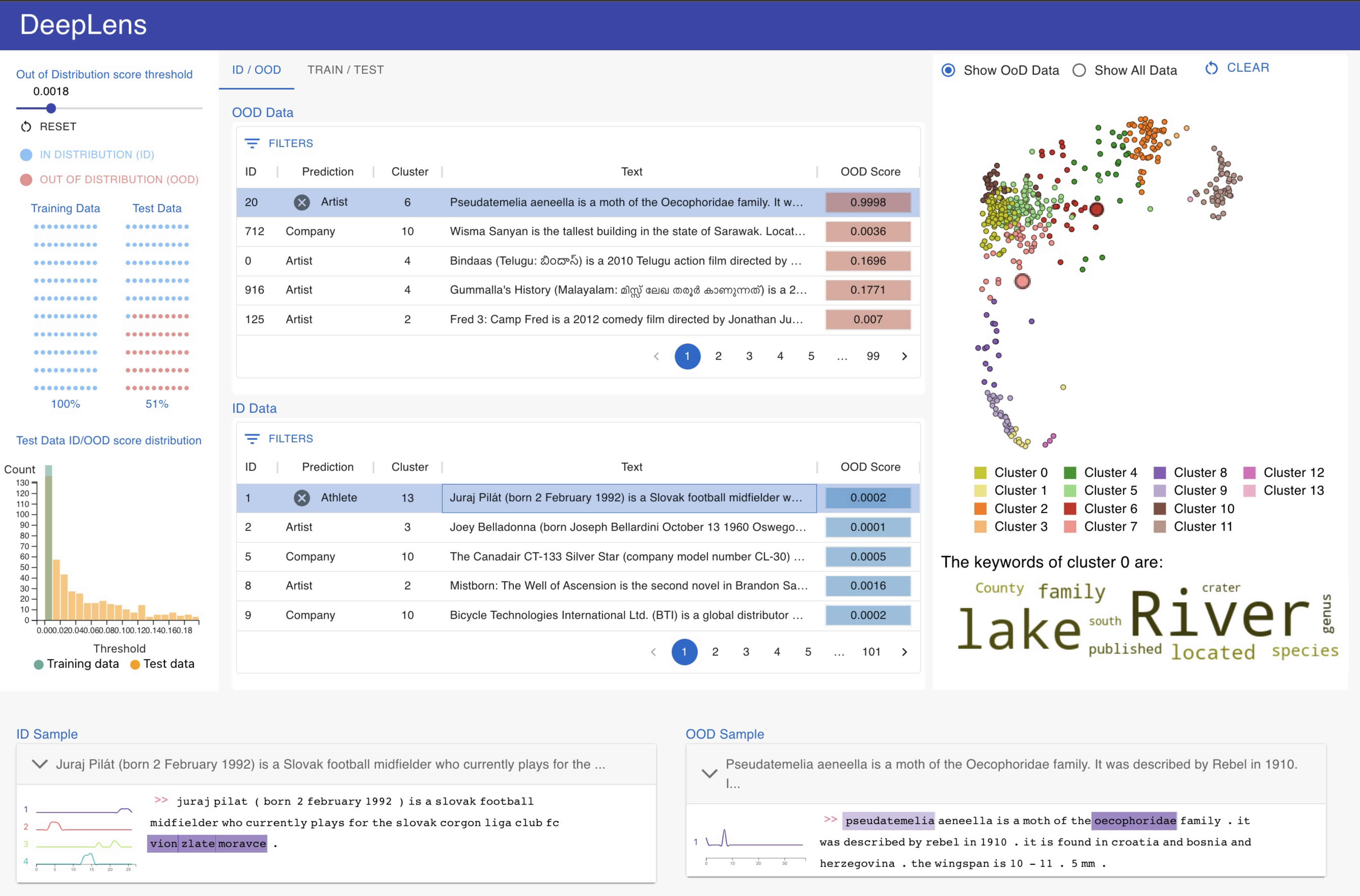}
  \caption{Interface of {\tool} for task 1.}
  \Description{This figure shows the interface of DeepLens for Topic Classification task.}
  \label{fig:interface_t1}
\end{figure*}

\subsection{NLP Task 2: Sentiment Analysis}

In this task, we use IMDB datasets as ID data and Yelp Polarity binary sentiment classification datasets as OOD data. The IMDB dataset contains movie reviews. The Yelp polarity dataset is formed by reviews for different businesses. Both IMDB and Yelp datasets have two labels Positive and Negative to predict the sentiment of the reviews. To create online data with background shift, we sample 1000 instances from both IMDB and Yelp test splits. 

\noindent\textbf{In-distribution Data}
\responseline{495 samples of }movie reviews from IMDB dataset.

\noindent\textbf{Out-of-distribution Data}
\responseline{505 samples of }business reviews from Yelp dataset.

\noindent\textbf{Model}
We fine-tune the BERT model for 1 epoch on the training split of ID data with a learning rate of $5\times10^{-5}$ and a batch size of 16. It achieves 93.5\% accuracy on the validation splits of ID data and 89\% accuracy on the \responseline{test} data.

\begin{figure*}[t]
  \centering
  \includegraphics[width=0.85\linewidth]{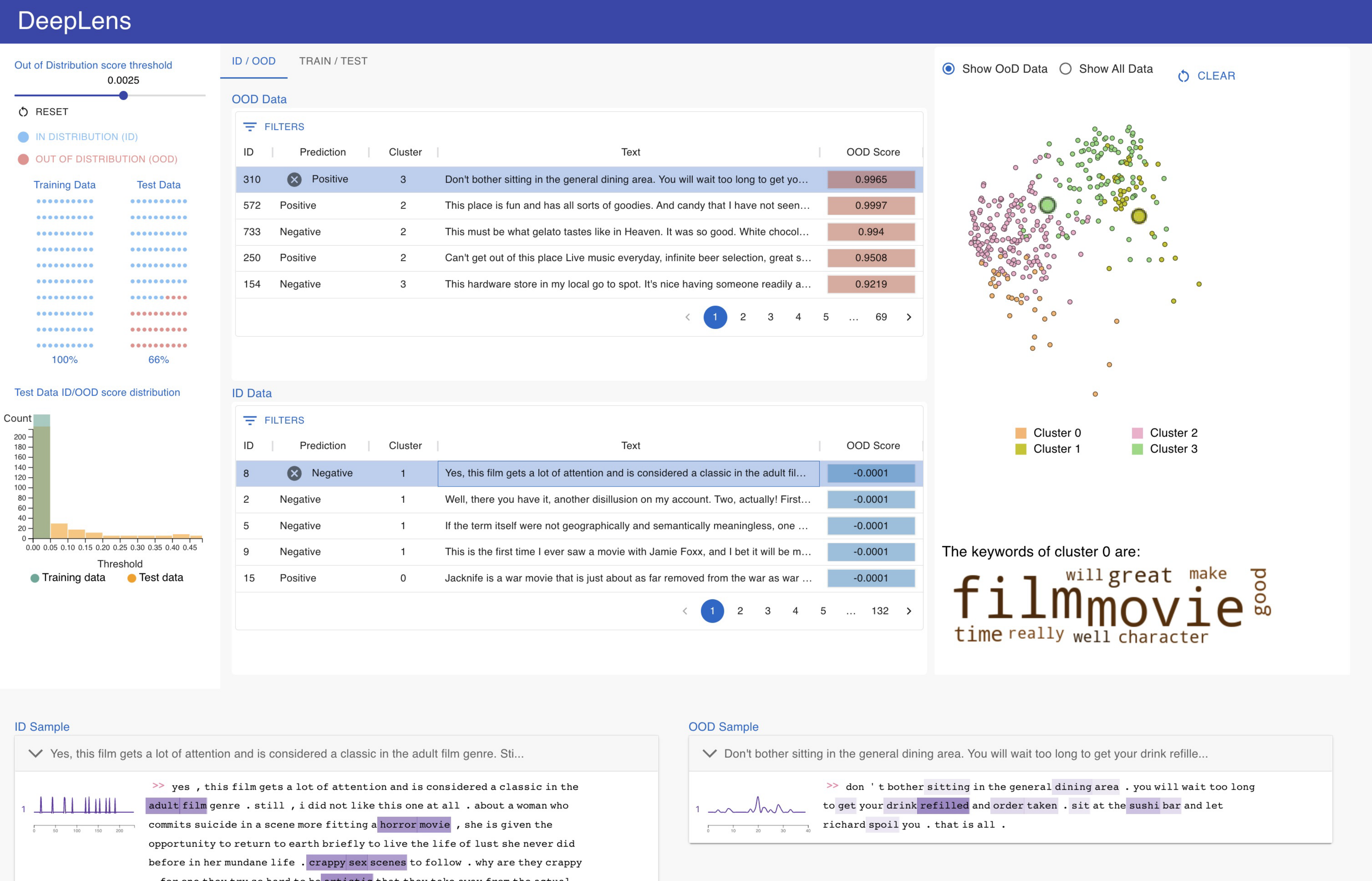}
  \caption{Interface of {\tool} for task 2.}
  \Description{This figure shows the interface of DeepLens for Sentiment Analysis task.}
  \label{fig:interface_t2}
\end{figure*}

\subsection{NLP Task 3: Emotion Recognition}

The six basic emotions included in the Emotion dataset are Sadness, Fear, Joy, Anger, Surprise, and Love. The source of the dataset is English Twitter Messages. There are 2 columns in the dataset, mapping to emotion index (0 to 5) and text. We use examples from the Sadness and Fear classes as ID data and the rest as OOD data. To create online data with semantic shift, we sample 1000 instances from 6 topics' test splits. 

\noindent\textbf{In-distribution Data}
ID data contains \responseline{644 samples and} 2 emotions: Sadness and Fear.

\noindent\textbf{Out-of-distribution Data}
OOD data contains \responseline{356 samples and} 4 emotions: Joy, Anger, Surprise, and Love.

\noindent\textbf{Model}
We fine-tune the BERT model for 4 epochs on the training split of ID data with a learning rate of $2\times10^{-5}$ and a batch size of 32. It achieves 99.2\% accuracy on the validation splits of ID data and 31\% accuracy on the \responseline{test} data.

\begin{figure*}[t]
  \centering
  \includegraphics[width=0.85\linewidth]{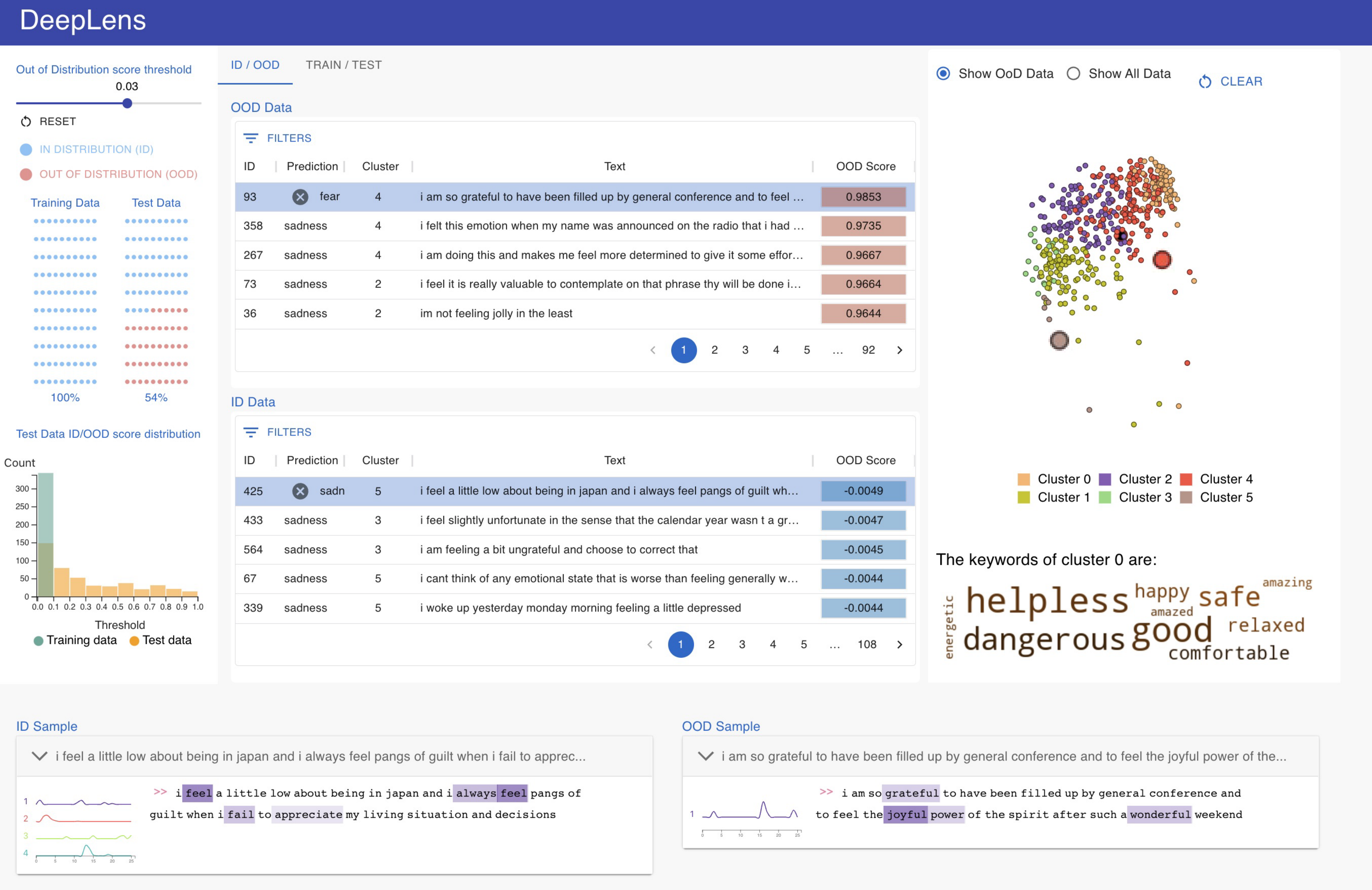}
  \caption{Interface of {\tool} for task 3.}
  \Description{This figure shows the interface of DeepLens for Emotion Recognition task.}
  \label{fig:interface_t3}
\end{figure*}

\subsection{NLP Task 4: Fake News Detection}

In this task, we design a background shift scenario that involves fakeness detection on different types of news. FakeNewsNet is a dataset collected from two fact-checking websites: GossipCop and PolitiFact. It contains news with labels indicating its validity annotated by professional journalists and experts. PolitiFact contains news related to U.S. politics and GossipCop is formed by entertainment news and gossip news. Besides, we also use the COVID-19 Fake News dataset in this task. It contains COVID-19-related news extracted from social media such as Facebook, Twitter, etc.  We use PolitiFact data as ID data. Then we combine GossipCop and COVID-19 Fake News datasets as OOD data. To create online data with background shift, we sample 4500 instances from PolitiFact, GossipCop, and COVID-19 Fake News test splits. 

\noindent\textbf{In-distribution Data}
\responseline{2000 samples of} news related to U.S. politics.

\noindent\textbf{Out-of-distribution Data}
\responseline{2500 samples of} news related to gossip and COVID-19. 

\noindent\textbf{Model}
We fine-tune the BERT model for 3 epochs on the training split of ID data with a learning rate of $5\times10^{-5}$ and a batch size of 16. It achieves 89.5\% accuracy on the validation splits of ID data and 67\% accuracy on the \responseline{test} data.

\begin{figure*}[t]
  \centering
  \includegraphics[width=0.85\linewidth]{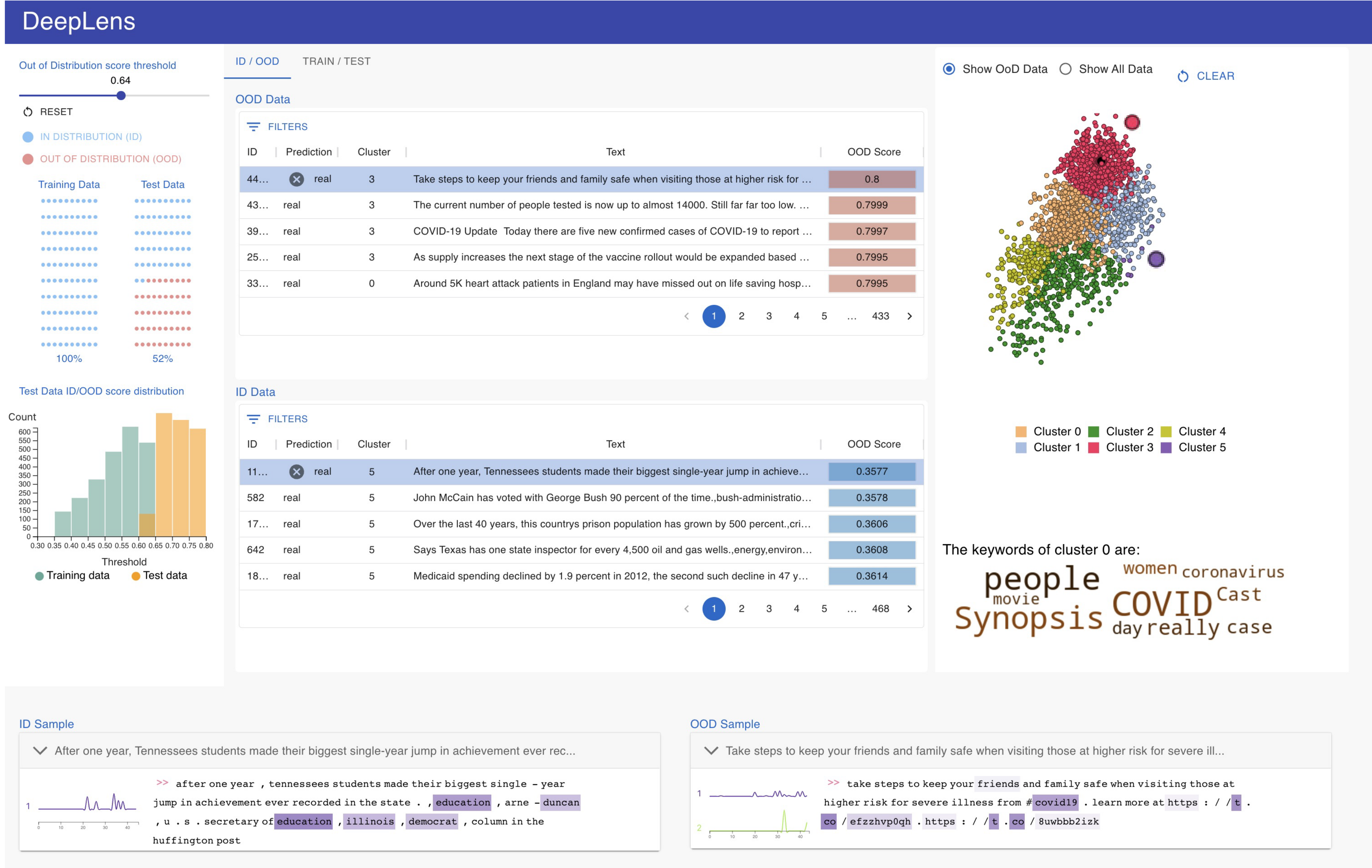}
  \caption{Interface of {\tool} for task 4.}
  \Description{This figure shows the interface of DeepLens for Fake News Detection task.}
  \label{fig:interface_t4}
\end{figure*}

\end{document}